\documentclass[1p]{elsarticle}
\usepackage{amssymb}
\usepackage{amsmath}
\usepackage{amsfonts}
\usepackage{algorithm, algorithmicx, algpseudocode}
\usepackage{pdfpages}
\usepackage{lineno}
\usepackage{enumitem}
\usepackage{mathtools}
\usepackage{graphicx}
\usepackage{caption} 
\usepackage{booktabs}

\usepackage{subcaption}
\usepackage{array,multirow}
\usepackage{amsthm}
\usepackage{threeparttable}
\usepackage{mathtools}
\usepackage{booktabs}
\usepackage{amsmath}
\usepackage{geometry}
\usepackage{longtable}
\usepackage{pdflscape}
\usepackage{tabularx}
\usepackage{multicol}
\usepackage{float}
\usepackage{adjustbox}
\usepackage{tablefootnote}
\usepackage{lipsum}
\usepackage{algorithmicx}

\usepackage{tikz}
\usepackage{pgfplots}
\pgfplotsset{compat=1.8} 
\usepackage{pgfplotstable}

\usepackage{xcolor, soul}
\sethlcolor{yellow}
\biboptions{sort&compress}
\usepackage[colorlinks,citecolor=blue,linkcolor=blue]{hyperref}
\makeatletter
\pretocmd{\NAT@open}{\begingroup\color{\@citecolor}}{}{}
\apptocmd{\NAT@close}{\endgroup}{}{}
\makeatother
\usepackage{xstring}
\makeatletter
\AtBeginDocument{
\let\oldref\ref
\renewcommand{\ref}[1]{\IfBeginWith{#1}{fig:}%
{{\color{blue}Fig.~\oldref{#1}}}%
{\IfBeginWith{#1}{lem:}{{\color{blue}Lemma~\oldref{#1}}}
{\IfBeginWith{#1}{tab:}{{\color{blue}Table~\oldref{#1}}}
{\IfBeginWith{#1}{sec:}{{\color{blue}Section~\oldref{#1}}}
{\IfBeginWith{#1}{eq:}{{\color{blue}Eq.~(\oldref{#1})}}
~}}}%
}}}
\makeatother
\usepackage{natbib}
\usepackage{cleveref}
\begin{document}
\begin{frontmatter}

\title{A Novel Pseudo-Random Number Generator Based on Multi-Objective Optimization  for Image-Cryptographic Applications}
 \author[aa]{Takreem Haider\corref{bb}}
 \ead{tahaider@iu.edu}
  \author[aa]{Sa\'ul A. Blanco}
 \ead{sblancor@indiana.edu}
 \author[ab]{Umar Hayat\corref{bb}}
 \ead{u.hayat@surrey.ac.uk} 
 \cortext[bb]{Corresponding author}
\address[aa]{Department of Computer Science, Indiana University Bloomington, IN 47408, USA}
 \address[ab]{Department of Computer Science, University of Surrey, Guildford, Surrey, GU2 7XH, UK}
 \begin{abstract}
Pseudo-random number generators (PRNGs) play an important role to ensure the security and confidentiality of image cryptographic algorithms. Their primary function is to generate a sequence of numbers that possesses unpredictability and randomness, which is crucial for the algorithms to work effectively and provide the desired level of security.
However, traditional PRNGs frequently encounter limitations like insufficient randomness, predictability, and vulnerability to cryptanalysis attacks.
To overcome these limitations, we propose a novel method namely an elliptic curve genetic algorithm (ECGA) for the construction of an image-dependent pseudo-random number generator (IDPRNG) that merges elliptic curves (ECs) and a multi-objective genetic algorithm (MOGA).
The ECGA consists of two primary stages. First, we generate an EC-based initial sequence of random numbers using pixels of a plain-image and parameters of an EC, that depart from traditional methods of population initialization.
In our proposed approach, the image itself serves as the seed for the initial population in the genetic algorithm optimization, taking into account the image-dependent nature of cryptographic applications. This allows the PRNG to adapt its behavior to the unique characteristics of the input image, leading to enhanced security and improved resistance against differential attacks.
Furthermore, the use of a good initial population reduces the number of generations required by a genetic algorithm, which results in decreased computational cost.
In the second stage, we use well-known operations of a genetic algorithm to optimize the generated sequence by maximizing a multi-objective fitness function that is based on both the information entropy and the period of the PRNG.
By combining elliptic curves and genetic algorithms, we enhance the randomness and security of the ECGA.
To evaluate the effectiveness and security of our generator, we conducted comprehensive experiments using various benchmark images and applied several standard tests, including the National Institute of Standards and Technology (NIST) test suite.
We then compared the results with the state-of-the-art PRNGs.
The experimental results demonstrate that the ECGA outperforms the state-of-the-art PRNGs in terms of uniformity, randomness, and cryptographic strength.
\end{abstract}
\end{frontmatter}

\section{Introduction}
Pseudo-random number generators (PRNGs) are extensively used in numerous fields, such as statistics, computer science, cryptography, and gaming~\cite{menezes2018handbook, luby1996pseudorandomness}. PRNGs generate a sequence of numbers that appear random but are produced from a predetermined starting point using a mathematical formula. The quality of a PRNG is of utmost importance in the field of cryptography as the security of a cryptographic system depends on the randomness and unpredictability of the keys generated. 
A good PRNG should possess the following key features to ensure the quality and security of the generated random 
numbers~\cite{knuth1998art}. 
\begin{enumerate}[itemsep=0pt]
\item[1)]
Randomness: To achieve the quality of a reliable PRNG, the generated number sequence must exhibit no distinguishable characteristics when compared to a truly random sequence. The generated output should be devoid of any identifiable patterns, and each number in the sequence should not correlate with the numbers that precede or follow it.
\item[2)]
Unpredictability: The PRNG needs to possess resistance against attacks aimed at forecasting future outputs by analyzing past outputs. To achieve this, the PRNG must possess a substantial internal state and employ a robust cryptographic algorithm.
\item[3)]
Periodicity: Each PRNG has a specific point at which the sequence it produces starts repeating. A PRNG is considered to be of high quality if its period is significantly longer, meaning it is approximately equal to the total number of possible outputs. 
\item[4)]
Security: The primary focus of the PRNG should be on maintaining strong security measures and guaranteeing resilience against commonly recognized forms of attacks, such as brute-force attacks.
\item[5)]
Efficiency: The PRNG should possess efficiency in terms of both computational speed and memory consumption, particularly for tasks that involve generating a substantial number of random values, such as the creation of cryptographic keys.
\end{enumerate}

Designing a PRNG with optimal randomness is a challenging task that requires balancing many different factors~\cite{ullah2021efficient}. The quality of random numbers generated by a PRNG is crucial in many applications, and it is essential to carefully evaluate and choose a PRNG that meets the quality and security requirements of the application or system. \\

In recent years, chaotic systems have become popular in the development of PRNGs due to their desirable features such as unpredictability, irreducibility, sensitivity to initial conditions, ergodicity, and chaoticity~\cite{thietart1995chaos}. 
Various PRNGs have been designed based on chaotic maps, for instance, PRNGs described 
in~\cite{murillo2017novel,hamza2017novel,xia2018novel,meranza2019pseudorandom,barani2020new,
zhao2019self,wang2019pseudo, gayoso2013pseudorandom,gayoso2019general, yu2021design, cang2021pseudo,
agarwal2021designing, shi2021hybrid, zang2022construction}.
Murillo et al.~\cite{murillo2017novel} introduced a PRNG that uses an improved 1D logistic map to generate pseudo-random numbers with strong statistical characteristics.
Hamza~\cite{hamza2017novel} presented a method that uses the Chen chaotic system to construct a PRNG for cryptographic purposes involving images. This method~\cite{hamza2017novel}  addresses the issue of non-uniform distribution commonly found in pseudo-random number sequence (PRNS) generated by the Chen chaotic system and produces PRNS with a high level of randomness.
Xia and Zheng~\cite{xia2018novel}, developed a novel PRNG that utilizes a controlled digital chaotic system. The purpose of this generator is to enhance the dynamic degradation that arises from the use of chaotic systems.
Meranza et al.~\cite{meranza2019pseudorandom} utilizes an improved version of the Henon map to design a PRNG. Their research indicates that the cryptographic properties of PRNS generated by the enhanced Henon map are superior to those produced by the traditional Henon map.
Barani et al.~\cite{barani2020new}, designed a PRNG for creating PRNS  by utilizing a generalized Newton complex map. To ensure the randomness of the generated sequences, several security measures were implemented and the outcomes indicated that this generator can produce secure PRNS.
Zhao et al.~\cite{zhao2019self} used a hyper-chaotic system to design a PRNG that exhibits high levels of randomness.
In addition, Wang et al.~\cite{wang2019pseudo}  constructed a PRNG that is based on a logistic chaotic system.
Gayoso et al.~\cite{gayoso2013pseudorandom} introduced a new PRNG that utilizes the residue number system, which enables the creation of an exceptionally efficient circuit that operates distinctly compared to conventional generators.
Furthermore, in~\cite{gayoso2019general}, the authors create a structure resembling a Hopfield neural network where each neuron is substituted with a compact PRNG. 
Yu et al.~\cite{yu2021design} developed a PRNG that uses a chaotic system and an improved Hopfield neural network. Their PRNG is designed to decrease the impact of chaotic degradation and enhance the quality of PRNS.
Cang et al.~\cite{cang2021pseudo}  presented a PRNG based on a generalized conservative Sprott-A chaotic system.
Agarwal et al.~\cite{agarwal2021designing} designed a PRNG that is based on the cascade fractal function. The cascade function is created using a combination of two seed maps, which improves the unpredictability and randomness of PRNG.
Shi and Deng~\cite{shi2021hybrid} proposed a new PRNG that is based on Baker chaotic map and can generate highly random PRNS.
Zang et al.~\cite{zang2022construction} developed an algorithm for generating PRNS using complex polynomial chaotic maps. The PRNS generated by this method shows strong randomness and are vulnerable to differential attacks.
A significant limitation associated with chaos-based cryptography arises from the fact that chaotic maps are designed to work with real numbers, which is not ideal for cryptographic applications that use finite numbers. 
Round-off errors in quantizing real numbers can create issues that result in irreversible functions, making decryption impracticable.\\

The use of elliptic curve cryptography (ECC) is gaining popularity in modern cryptographic applications due to its effectiveness, strong security measures, and resilience against attacks. 
The difficulty of solving the elliptic curve discrete logarithm problem (ECDLP) is a significant factor that motivates the preference for elliptic curves (ECs) over chaotic maps in the design of cryptographic algorithms.
Furthermore, ECs exhibit the advantage of necessitating significantly reduced key sizes in comparison to chaotic maps. This characteristic renders them more efficient and viable for implementation within environments that face limitations in terms of available resources.
As a result, various PRNGs using the arithmetic of ECs have been designed.
Hayat and Azam~\cite{hayat2019novel}, proposed an algorithm for generating PRNS which is based on ordered ECs. 
This method is efficient when compared with previously introduced PRNGs over ECs.
However, the generator~\cite{hayat2019novel} is not suitable for ECs over large prime $p$ due to 
high space and time complexity such as $\mathcal{O}(p)$ and $\mathcal{O}(p^{2})$, respectively.
A PRNG based on Mordell elliptic curve is introduced by Ullah et al.~\cite{ullah2021efficient}, which is more efficient and has better cryptographic properties than~\cite{hayat2019novel}.
However, the time and space complexity of the generator~\cite{ullah2021efficient} is $\mathcal{O}(mp)$  and $\mathcal{O}(m)$, respectively, where  $m \leq p$ is the size of PRNS, due to which this generator is not compatible with ECs associated with large prime $p$. 
An isomorphic EC-based PRNG, developed by Haider et al.~\cite{haider2022novel}, produces sequences with high randomness and outperforms existing generators in terms of cryptographic properties; however, it faces compatibility issues with large prime ECs when the parameters of EC and the size of the ordered set are not predetermined. 
Recently, Adhikari and Karforma~\cite{adhikari2022novel} presented a PRNG over large prime ECs. To generate pseudo-random numbers, firstly, the $y$-coordinate of the generated point over the EC is extracted and then the least significant 8 bits of the extracted $y$-coordinate are converted to its decimal representation.
Although this PRNG is compatible with ECs over large primes to obtain PRNS with length $\ell$, this algorithm needs to generate $\ell$ number of points over the EC. So, for large $\ell$, this method is not suitable for real-world applications.
The existing EC-based PRNGs have exhibited favorable outcomes, however, they do not guarantee the generation of PRNs with security levels closely approximating the theoretically optimal values.\\
\subsection{Our contribution}

To address the aforementioned issues, our focus is directed toward the design of a PRNG that produces random numbers of high quality, exhibiting optimal randomness. The following steps outline our contributions:
\begin{enumerate}[itemsep=0pt]
\item[1)] To overcome the challenges posed by low randomness, predictability, and vulnerability to cryptanalysis attacks, we employ a multi-objective genetic algorithm (MOGA) optimization technique. This approach is chosen because MOGA allows us to simultaneously optimize multiple objectives, such as randomness, predictability, and resistance to cryptanalysis attacks. By employing MOGA, we can enhance the overall quality and security of the generated random numbers.
\item[2)] Our method takes advantage of the image itself as the seed for the initial population in the genetic algorithm optimization process. This choice is motivated by the image-dependent nature of cryptographic applications. By using the image as the seed, the PRNG can adapt its behavior to the unique characteristics of the input image. This adaptation improves the security of the PRNG, making it more resistant to differential attacks and increasing the level of protection against potential threats.
\item[3)] We generate an initial sequence of random numbers based on both the plain-image and elliptic curve. This departure from traditional methods of population initialization is selected for a specific reason. By using the plain-image and elliptic curve, we ensure that the initial solution provided to the genetic algorithm is well-chosen and has desirable properties. This decreases the number of generations required by the genetic algorithm and thus minimizes the overall computation time.
\item[4)] The genetic algorithm is utilized to improve the generated sequence by maximizing a fitness function that considers both information entropy and the period of the pseudo-random sequence. This choice is made because the genetic algorithm is effective in optimizing problems that have multiple objectives. By maximizing the fitness function, we can enhance the information entropy and period of the generated sequence, leading to superior-quality random numbers.
\item[5)] To evaluate the performance and security of our proposed PRNG, extensive experiments are conducted using various benchmark images. The results are then compared with the existing state-of-the-art PRNGs. The experiments serve as empirical evidence supporting our claims of enhanced performance and security.
\end{enumerate}

\subsection{Paper organization}

The rest of the paper is organized as follows. 
In~\ref{sec:preliminaries}, we present the preliminary theoretical background and notions used in the paper. 
Moreover, in~\ref{sec:generator}, we describe our method ECGA for the construction of the proposed IDPRNG based on elliptic curves and a genetic algorithm.
In~\ref{sec:SA_C}, we present a comprehensive security analysis and the comparison of the ECGA with the 
state-of-the-art PRNGs, establishing empirically that our method is superior. 
Finally, in~\ref{sec:Con}, we summarize the findings of the ECGA and discuss possible future directions of our line of work.

\section{Preliminaries}\label{sec:preliminaries}
In this section, we present fundamental concepts that are used in our algorithm and hold significance for comprehension. 
\subsection{Elliptic Curves}
Let $p>3$ be a prime number and $a,b $ be two integers, and we use $\mathbb{F}_p$ to denote the finite field of $p$ elements. 
  An elliptic curve (EC) denoted by $E_{p,a,b}$ is defined as a set of solutions $(x,y) \in \mathbb{F}_p \times \mathbb{F}_p$ satisfying the equation $y^{2} \equiv x^{3}+ ax +b \pmod{p}$, along with an additional point at infinity denoted by $\infty$, where $a, b \in \mathbb{F}_{p}$ and $4a^{3}+27b^{2}\not\equiv 0 \pmod{p}$.\\

The group law operation $+_{g}$ on an elliptic curve $E_{p,a,b}$ is defined as follows~\cite{silverman2009arithmetic}. We denote the multiplicative inverse of $a\in \mathbb{F}_p$ by $a^{-1}$.\\

\noindent
\textbf{Point addition:}\\
Let $G_{1} = (x_{1},y_{1})$ ans $ G_{2} = (x_{2},y_{2})$ be two points on  $ E_{p,a,b}$ such that $G_{1} \neq G_{2}$ and 
$x_{1} \neq x_{2}$, then 
the computation of the resulting point $G_{1} +_{g} G_{2} = (x_3, y_3)$ when performing point addition over $ E_{p,a,b}$ is given as:
\begin{align*}
\lambda &\equiv (y_{2} - y_{1})(x_{2} - x_{1})^{-1} \pmod{p}.\\
x_{3} &\equiv \lambda^{2} - x_{1} - x_{2} \pmod{p},\\
y_{3} &\equiv \lambda(x_{1} - x_{3}) - y_{1} \pmod{p}.
\end{align*}

\noindent
\textbf{Point doubling:}\\
Let $G_{1} = (x_{1},y_{1})$  be a point on  $ E_{p,a,b}$ such that $y_{1} \neq 0$, then 
the computation of the resulting point $G_{1} +_{g} G_{1}=(x_3, y_3)$ when performing point doubling over $ E_{p,a,b}$ is given as:
\begin{align*}
\lambda &\equiv (3 x^{2}_{1})(2 y_{1})^{-1} \pmod{p}\\
x_{3} &\equiv \lambda^{2} - 2x_{1}   \pmod{p}, \\
y_{3} &\equiv \lambda(x_{1} - x_{3}) - y_{1} \pmod{p}.
\end{align*}


\subsection{Genetic Algorithm}
In recent times, there has been significant interest among researchers in \emph{evolutionary algorithms} (EAs), which have been recognized as valuable in numerous applications (see, e.g.,~\cite{abdullah2012hybrid, haider2022novel} and references within).  One of the most widely recognized types of EAs is  \emph{genetic algorithms} (GAs), 
which are search heuristics. Several applications have been devised using genetic algorithms~\cite{abdullah2012hybrid, wu2006genetic}. Fundamentally, a GA can be divided into four primary stages~\cite{zames1981genetic}, which we describe below.

\begin{description}
    \item{Initial population:}
During the creation of the population, an initial set of individuals or solutions is generated. These individuals are designed to form the starting point for the genetic algorithm. They represent possible solutions to the problem being addressed and are encoded in a format that allows the algorithm to work with and manipulate them.
    \item{Selection:} During the selection step, a portion of individuals from the population is selected as parents for the next generation. This selection is primarily determined by the fitness or quality of each individual solution. Individuals with higher fitness scores are more likely to be chosen as parents since they are considered to have superior solutions.
\item{Crossover:} During the crossover step, new solutions are generated by merging the genetic material of chosen parent individuals. The genetic material, typically represented as chromosomes or sequences of values, is swapped between parents to produce offspring. This procedure emulates the natural recombination of genes that takes place during reproduction.
\item{Mutation:} Mutation involves a spontaneous and random alteration in a specific aspect or trait of a solution. During the mutation step, slight modifications are made to the genetic composition of individual solutions. This randomness plays a crucial role in introducing fresh genetic variations within the population. 
\end{description}

A genetic algorithm explores the search space through multiple iterations, aiming to discover an optimal or nearly optimal solution for the given problem.

\section{Elliptic Curve Genetic Algorithm (ECGA)} \label{sec:generator}
In this section, we provide a novel method namely elliptic curve genetic algorithm (ECGA) for generating pseudo-random numbers by using elliptic curves and a genetic algorithm. The effectiveness of a PRNG highly depends on its tendency to produce random and unpredictable sequences.
The quality of a PRNG is significantly enhanced by two important factors, namely high entropy and a long period.
The presence of high entropy in a sequence of numbers makes it difficult for an attacker to predict the next number, while a long period decreases the chances of repetitive patterns that could be used for exploitation.
We propose a generator that consists of two important stages. Initially, we utilize points on elliptic curves to create a sequence of random numbers.
Subsequently, we employ the operations of a genetic algorithm to maximize a fitness function that considers multiple objectives. This fitness function guarantees a higher degree of randomness by considering both the information entropy and the period of the sequence.
By employing an appropriate initial solution based on ECs, the optimization process reduces the number of generations needed, thus decreasing computational time.
Furthermore, our proposed method ECGA enhances the unpredictability and randomness of the PRNG by incorporating elliptic curves and a genetic algorithm.  
We provide a concise overview of each stage of the ECGA as follows.

\subsection{Initialization}
We use elliptic curves to generate an initial sequence of pseudo-random numbers. 
The initialization process comprises the following procedures:

\noindent 
\textbf{Step 1:}
Let $I$ be a plain-image in a two-dimensional array with dimensions $r\times s$, where each element belongs to the symbol set $[0, 2^{m}-1]$  and $m$ represents the number of bits in the pixel of an image. 
Here, if $n_1,n_2$ are two integers with $n_1<n_2$, $[n_1,n_2]=\{n_1,n_1+1,\ldots,n_2\}$. 
We consider an elliptic curve $E_{p,a,b}$ defined by the equation $y^{2} \equiv x^{3}+ax +b \pmod{p}$, where $p$, $a$, and $b$ are parameters of the curve and $G=(x_{0},y_{0})$ is a base point on the curve. 
Let us assume that $H_{I}$, $H_{a}$, $H_{b}$, and $H_{p}$ denote the SHA-256 hash value of $I$, $a$, $b$, and $p$, respectively.  
The proposed approach begins by selecting an initial point denoted as $G_{0}=(x_{0},y_{0})$ on  $E_{p, a,b}$. Subsequently, a series of $n$ points $G_{k}=(x_{k},y_{k})$ are generated, where $1 \leq k \leq n$. 

\noindent 
\textbf{Step 2:} 
Calculate the binary values for the $x$ and $y$ coordinates of the point $G_{k}=(x_{k},y_{k})$, with $1\leq k\leq n$.
We define the function $B(x_{k})$ that takes as input the decimal representation of $x_{k}$ and outputs its binary representation as a sequence of $u$ bits: $B(x_{k}) =x_{k}^{1},x_{k}^{2}, \ldots , x_{k}^{u}$, where $u$ is the number of bits needed to represent $x_{k}$ in binary form. Similarly, we define the function $B(y_{k})$ that takes as input the decimal representation of $y_{k}$ and outputs its binary representation as a sequence of $v$ bits: $B(y_{k}) =y_{k}^{1},y_{k}^{2}, \ldots , y_{k}^{v}$, where $v$ is the number of bits needed to represent $y_{k}$ in binary form.

\noindent 
\textbf{Step 3:} 
From Step 1, let $\jmath \in \lbrace I, p, a, b \rbrace$ and $H_\jmath$ be the SHA-256 sets from Step 1. Namely, 
$H_{\jmath} = \lbrace h_{\jmath}^{i} : 1 \leq i \leq 256, h_{\jmath}^{i} \in \lbrace 0, 1 \rbrace \rbrace.$ 
We furthermore define $B^{x}_{a,b}$ to be a binary sequence of length $3 \ell'$, where  $\ell'= \text{min}(u,256)$, obtained by repeatedly merging the corresponding binary bits of $H_{a}$, $B(x_{k})$, and $H_{b}$ in an alternating manner as follows:
\begin{equation}
B^{x}_{a,b} = h_{a}^{1}, x_{k}^{1}, h_{b}^{1},  h_{a}^{2}, x_{k}^{2}, h_{b}^{2}, \ldots,  h_{a}^{\ell'},x_{k}^{\ell'},h_{b}^{\ell'}.
\end{equation}
Similarly, we define $B^{y}_{I,p}$ to be a binary sequence of length $3 \ell''$, where  $\ell''= \text{min}(v,256)$, obtained by merging the binary bits of $H_{I}$, $B(y_{k})$, and $H_{p}$ as follows:
\begin{equation}
B^{y}_{I,p} = h_{I}^{1}, y_{k}^{1}, h_{p}^{1},  h_{I}^{2}, y_{k}^{2}, h_{p}^{2}, \ldots,  h_{I}^{\ell''},y_{k}^{\ell''},h_{p}^{\ell''}.
\end{equation}
\noindent 
\textbf{Step 4:} 
Generate a  binary sequence $B^{x,y}_{I, p, a,b}$ of length $\ell$, where $\ell = 3(\ell' + \ell'')$, by concatenating $B^{x}_{a,b}$ and $B^{y}_{I,p}$ as follows:
\begin{equation}
B^{x,y}_{I, p, a,b} = B^{x}_{a,b} +_{\rm{C}} B^{y}_{I,p}.
\end{equation}
Note that the concatenation operation denoted by $+_{\rm{C}}$ is simply the operation of appending the second sequence to the end of the first sequence.\\
\noindent 
\textbf{Step 5:} 
Let $B_z$ denote a randomly selected binary sequence of length $\ell$.  We define the resultant binary sequence $B^{x,y}_{I, p, a,b,z}$ of length $\ell$ as follows: 
\begin{equation}
B^{x,y}_{I, p, a,b,z}=\begin{cases}
0 & \text{if } \xi_i=\xi'_i~\forall i\in\lbrace 1,2,\ldots,\ell \rbrace \\
1 & \text{if } \xi_i\neq\xi'_i~\forall i\in\lbrace 1,2,\ldots,\ell\rbrace
\end{cases},
\end{equation}
where $\xi_{i}$ and $\xi'_{i}$ are the $i$-th element of  the binary sequences $B^{x,y}_{I, p, a,b}$ and $B_z$, respectively.\\
\noindent 
\textbf{Step 6:} 
Divide the binary sequence $B^{x,y}_{I, p, a,b,z}$ into segments of a predetermined length, $m$, and convert each segment into its decimal representation. Let $S_{d}$ denotes the decimal representation of $B^{x,y}_{I, p, a,b,z}$. Therefore, $S_{d}$ can be expressed as a sequence of decimal values $\delta_{1}, \delta_{2}, ..., \delta_{t}$, where $t = \lfloor \ell/m \rfloor$ .
Note that each of the decimal values in the sequence $S_{d}$  falls within the range of $[0, 2^{m}-1]$.\\
\noindent 
\textbf{Step 7:} 
Repeat Step 2 through Step 6 for each $k \in \lbrace 1, 2, \ldots, n \rbrace$. 
Let $ \Delta(I,p,a,b,x,y,z,n)$ be the resulting sequence of length $n \times t$.\\
\noindent 
\textbf{Step 8:} 
Choose three positive integers $\phi$, $\psi$, and $\varphi$. The proposed IDPRNG is represented by the function
\begin{equation*}
\Omega : \Delta \rightarrow [0, 2^{m}-1], \quad \text{where}
\end{equation*}
 \begin{equation}
 \Omega(\delta_{i}) \equiv \phi \delta_{i} + \psi \delta_{i+1} + \varphi \pmod{2^{m}}.
 \end{equation}
 
Thus, $ \Omega(I,p, a,b,x,y,z,n,\phi, \psi, \varphi)$ is the initial sequence of random numbers based on the parameters 
$I,p,a,b,x,y,z,$ $n,\phi, \psi$, and $\varphi$. Henceforth, we represent it by $\Omega_{I}$.


\subsection{Fitness function}
A fitness function is a tool used to measure how closely a given solution matches the ideal solution. 
The proposed algorithm aims to find the best possible solution for a given problem. 
The degree of uncertainty in a PRNS is often measured using information entropy ($H$), which is an important measure of randomness. For a PRNS to be considered effective, it must contain a high level of uncertainty. The higher the entropy value, the stronger the generator is considered to be. 

\noindent
Let $\Omega $ be a PRNS taking values from $[0, 2^{m}-1]$.
The entropy $H(\Omega) $ of $\Omega $ is defined as:
\begin{equation}
H(\Omega) = - \sum_{i=1}^{2^{m}} \rm{P}(\omega_i) \text{log}_2 \rm{P}(\omega_i),
\end{equation}
where $\rm{P}(\omega_i)$ represents the probability of an  $i$-th element $\omega_i$ in $\Omega$.

\noindent
Apart from entropy, the period of a PRNS is also a significant factor in assessing its randomness. A PRNS with a long period is generally considered good for cryptographic purposes. The \emph{period} of $\Omega$, denoted by $T(\Omega)$, is the the least positive integer $T$ for which $\omega_{i + T} = \omega_{i}$ for all $i \geq 1$. The case $T(\Omega)=\ell(\Omega)$ is optimal since then $\Omega$ is considered more secure. 


\noindent
To achieve our objective, a multi-objective optimization function is employed that seeks to maximize both the information entropy and the period of a pseudo-random number sequence $\Omega$ of length $\ell$ which is initially generated. 
The purpose of this function is to generate PRNSs that have both maximum entropy and period, to obtain the best possible results.\\

\noindent
Our optimization problem is based on the following fitness function.\\
Maximize 
\begin{equation}
f(\Omega) = H(\Omega) + T(\Omega),
\end{equation}
where $0 \leq H(\Omega) \leq m $ and $1 \leq T(\Omega) \leq \ell $.


\subsection{Crossover}
Let $\Omega_{I} = (\omega_0, \omega_1, ..., \omega_{2^m-1})$ be the set of pseudo-random numbers generated during the initialization phase. The goal of the crossover operator is to replace elements in $\Omega_{I}$ with those that result in a higher fitness value. In other words, elements with lower fitness values are replaced with those having comparatively higher fitness values. More concretely, the crossover operation is carried out as follows:
\begin{enumerate}[itemsep=0pt]
\item[i)] {Let $P_{e}$ be a random permutation of the integers in the range $[1, 2^m]$.}
\item [ii)]{Let $V = (v_{1}, v_{2}, ... , v_{2^{m}})$ be a vector of $2^{m}$ elements randomly selected from $\Omega_{I}$.}
\item [iii)]{Define $\Omega_C$ as the sequence obtained by replacing the $v_i$-th element of $\Omega_{I}$ with the $i$-th element of $P_{e}$, i.e.,
\begin{equation}
\omega_{i}^{C}=\begin{cases}
\omega_{P_{e}(i)} & \text{if}~i = v_{j}~~\text{for some j} \\
\omega_{i} & \text{otherwise}
\end{cases}.
\end{equation}
}
\end{enumerate}
The purpose of this operation is to increase the diversity of the sequence and improve the chances of finding a globally optimal solution. Specifically, the crossover operation ensures that all integers in the range $[0, 2^{m}-1]$ are present in $\Omega_{C}$.

To evaluate the quality of the new sequence $\Omega_C$, we compute its entropy $H(\Omega_C)$ and period $T(\Omega_C)$. If $H(\Omega_{C}) \geq H(\Omega_{I})$ and $T(\Omega_{C}) \geq T(\Omega_{I})$, then we consider $\Omega_{C}$ as input for the mutation operation. Otherwise, if $H(\Omega_{C}) < H(\Omega_{I})$ or $T(\Omega_{C}) < T(\Omega_{I})$, we use $\Omega_{I}$ as input to the mutation phase.

\subsection{Mutation}
Let $\Omega_{M}$ be a sequence of length $\ell$ obtained after crossover operation, and let $r$ and $r'$ be two integers randomly selected from the interval $[1,\ell]$. The swapping mutation operator $\mu_s$ can be defined as:
\begin{equation}
\mu_{s}(\Omega_{M},r,r') = \Omega_{M'},
\end{equation}
where $\Omega_{M'}$ is the sequence obtained after swapping the element at position $r$ with the element at position $r'$ in $\Omega_{M}$.

Since the entropy of a sequence is not affected by the swapping operation, therefore in the mutation phase we only compute the period of the obtained sequence.
If $T(\Omega_{M'}) \geq T(\Omega_{M})$, then $\Omega_{M'}$ is selected for the next generation, otherwise, $\Omega_{M}$ is retained for the next generation.
\subsection{Termination }
The stopping criteria of any optimization algorithm depend primarily on two significant factors. The first factor is the number of generations or iterations, where if an algorithm attains the pre-defined number of generations, it terminates. The second factor is based on the optimal solution, where an algorithm terminates if it achieves the optimal solution to the fitness function. It is essential to note that the number of generations alone cannot serve as a feasible parameter to stop an algorithm since it may terminate without any improvement. As the objective is to generate highly random pseudo-random number sequences, therefore the proposed algorithm employs the optimal solution to the problem as the termination condition. In other words, the algorithm stops when the required sequence attains optimal entropy and optimal period, producing a highly random sequence that is well-suited for cryptographic applications.

Thus, the PRNS obtained after the termination phase represents the optimized sequence of pseudo-random numbers characterized by optimal entropy and period. We denote the optimized PRNS by $\Omega_{Z}$.


\section{Security analysis and comparison}\label{sec:SA_C}
This section evaluates and compares the effectiveness of  the ECGA through extensive experiments and security evaluations, which involve a variety of tests such as randomness analysis, entropy analysis, period analysis, Hurst exponent analysis, correlation analysis, key sensitivity analysis, and key space analysis. Additionally, we assess the efficiency of the ECGA in two different ways: 
\begin{enumerate}[itemsep=0pt]
\item[1)] by comparing the initially generated sequences with their optimized sequences, and 
\item[2)] by comparing the ECGA with the state-of-the-art 
generators~\cite{haider2022novel,zang2022construction,talhaoui2021new, paul2021design,cang2021pseudo, yu2021design, agarwal2021designing, shi2021hybrid, tutueva2020adaptive,barani2020new, ayubi2020deterministic, zhao2019self,wang2019pseudo, meranza2019pseudorandom,gayoso2019general,xia2018novel,murillo2017novel, hamza2017novel,gayoso2013pseudorandom}.
\end{enumerate}

For our experiments, we used MATLAB R2022b on a machine with an Intel Core m3-7Y30 @1.61 GHz and 8 GB of RAM. We generated 100 random sequences by selecting parameters at random, and the selected parameters are listed in~\ref{tab:parameters}. We used two elliptic curves recommended by NIST~\cite{qu1999sec}, namely $E_{p,a,b}$ and $E_{p',a',b'}$ with prime numbers of 256 bits and 521 bits, respectively. The parameters for these elliptic curves are provided in~\ref{tab:parameters}. In addition, we used four different standard images with three distinct dimensions and five sets of integer-based triplets to generate these sequences. Our approach generates these 100 sequences by modifying one of the four parameters, namely (a) the elliptic curve, (b) the plain-image, (c) the size of the plain-image, and (d) the triplet $(\phi, \psi, \varphi)$ while keeping the others constant.

\subsection{Randomness analysis}
\begin{table*}[t!]  
      \caption{The parameters used to generate pseudo-random sequences using the ECGA.}
        \label{tab:parameters}
   \begin{tabular*}{\textwidth}{l @{\extracolsep{\fill}}}
 \toprule  
1.~Elliptic curve  \\  \midrule
\textbf{$E_{p,a,b}:$}\\
$p=$~115792089210356248762697446949407573530086143415290314195533631308867097853951,\\
$a=$~115792089210356248762697446949407573530086143415290314195533631308867097853948,\\
$b=$~41058363725152142129326129780047268409114441015993725554835256314039467401291, \\
$G_{x} =$~48439561293906451759052585252797914202762949526041747995844080717082404635286,\\
$G_{y} =$~36134250956749795798585127919587881956611106672985015071877198253568414405109,\\
\textbf{$E_{p',a',b'}:$}\\
$p' = $~6864797660130609714981900799081393217269435300143305409394463459185543183397656052122
\\~~~~~~~559640661454554977296311391480858037121987999716643812574028291115057151,\\
$a' =$~6864797660130609714981900799081393217269435300143305409394463459185543183397656052122
\\~~~~~~~559640661454554977296311391480858037121987999716643812574028291115057148,\\
$b' =$~1093849038073734274511112390766805569936207598951683748994586394495953116150735016013
\\~~~~~~~708737573759623248592132296706313309438452531591012912142327488478985984,\\
$G_{x}' =$ 266174080205021706322876871672336096072985916875697314770667136841880294499642780849
\\~~~~~~~~1545080627771902352094241225065558662157113545570916814161637315895999846,\\
$G_{y}' =$ 375718002577002046354550722449118360359445513476976248669456777961554447744055631669
\\~~~~~~~~1234405012945539562144444537289428522585666729196580810124344277578376784.\\
\midrule
2.~Plain-image $I$  \\  
~~~
\begin{subfigure}[b]{.20\linewidth}
\centering
\includegraphics[width=\linewidth]{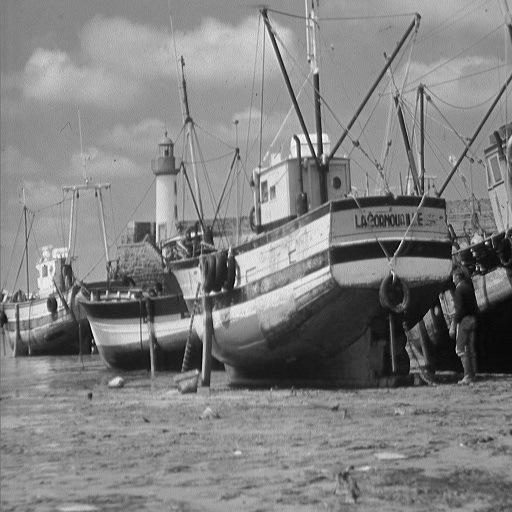}
\caption{ }\label{fig:a}
\end{subfigure}%
~~~
\begin{subfigure}[b]{.20\linewidth}
\centering
\includegraphics[width=\linewidth]{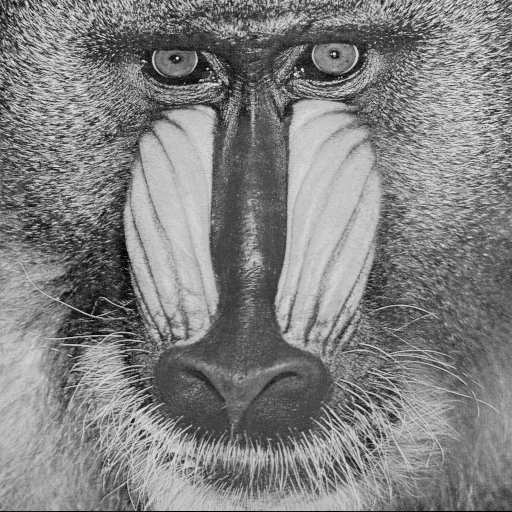}
\caption{ }\label{fig:b}
\end{subfigure}
~~~
\begin{subfigure}[b]{.20\linewidth}
\centering
\includegraphics[width=\linewidth]{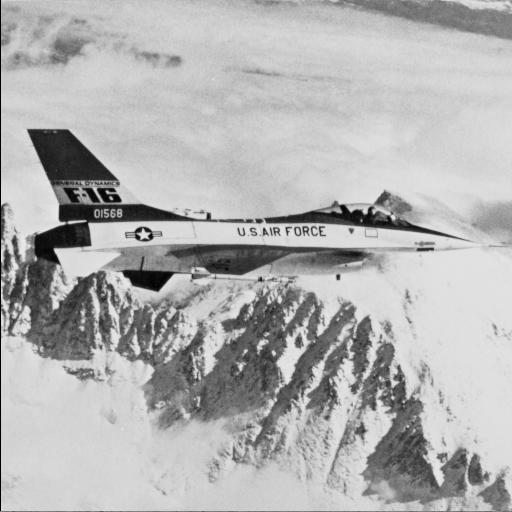}
\caption{ }\label{fig:c}
\end{subfigure}
~~~
\begin{subfigure}[b]{.20\linewidth}
\centering
\includegraphics[width=\linewidth]{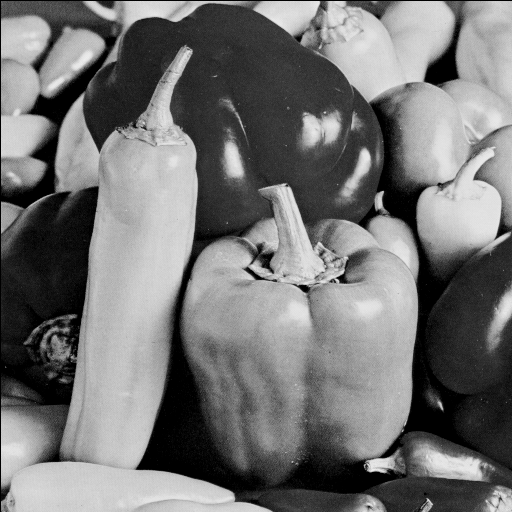}
\caption{ }\label{fig:d}
\end{subfigure}

\\ \midrule
3.~Size of  $I$\\ 
~~~$(r\times s) \in \lbrace (256 \times 256), (512 \times 512), (1024 \times 1024) \rbrace$. \\
\midrule
4.~The integers  $(\phi, \psi, \varphi)$  \\
~~~$(\phi, \psi, \varphi) \in \lbrace (25,73,121), (123,33,77), (785,221,9), (3,5,11), (17, 81, 955) \rbrace$.\\
 \bottomrule
\end{tabular*}
 \end{table*}
 \begin{table*}[!t]
      \caption{Parameters of NIST tests.}
        \label{tab:NIST_parameters}
   \begin{tabular*}{\textwidth}{l @{\extracolsep{\fill}}  lcll}
       \toprule   
No.	&	Test Name	&	Number	&	Recommended	&	\footnotemark[1]Parameter $M$ or $m$	\\
	&		&	of	sub-sets &~~~size $(n)$	&		\\ \midrule
1.	&	Frequency Monobit (FT)	&		&	$\geq 10^{2}  $ 	&		\\
2.	&	Block Frequency   (BFT)      	&		&	$\geq 10^{2}  $ 	&	$20 \leq  M \leq (n/100)$ 	\\
3.	&	Cumulative Sums (CST)	&	2	&	$\geq 10^{2}  $ 	&		\\
4.	&	Runs    (RT)	&		&	$\geq 10^{2}  $ 	&		\\
5.	&	Longest Runs    (LRT)	&		&	$\geq 128  $ 	&		\\
6.	&	Rank    (RRT)	&		&	$\geq 38912  $ 	&		\\
7.	&	Discrete Fourier Transform      (DFT)	&		&	$\geq 10^{3}  $ 	&		\\
8.	&	Non-overlapping Template        (NTMT)	&	148	&	$\geq 8m-8  $ 	&	$2 \leq  m \leq 21$ 	\\
9.	&	Overlapping Template    (OTMT)	&		&	$\geq 10^{6}  $ 	&		\\
10.	&	Universal Test  (UT)	&		&	$\geq 387840  $ 	&		\\
11.	&	Approximate Entropy     (AET)	&		&		&	 $ m < \lfloor \text{log}_{2}n \rfloor -5$ 	\\
12.	&	Random Excursions (RET)	&	8	&	$\geq 10^{6}  $ 	&		\\
13.	&	Random Excursions Variants (REVT)	&	18	&	$\geq 10^{6}  $ 	&		\\
14.	&	Serial      (ST)	&	2	&		&	 $2 < m < \lfloor \text{log}_{2}n \rfloor -2$ 	\\
15.	&	Linear Complexity    (LCT)   	&		&	$\geq 10^{6}  $ 	&	$500 \leq  M \leq 5000$ 	\\ 
  \bottomrule
\end{tabular*}
\\
\footnotesize{$^1$Some tests use a second parameter, which can be symbolized as either $m$ or $M$, and these parameters indicate the block length.}
\end{table*}
The NIST 800-22 test suite~\cite{rukhin2001statistical} is widely recognized as a suitable tool for evaluating the randomness of binary sequences. The suite consists of 15 tests and 174 sub-tests and requires, in general, at least 1 million bits to assess the randomness of a sequence. The suite calculates the probability of $p_{\text{value}}$ for each sequence, and if $p_{\text{value}} \geq \lambda$ (or $p_{\text{value}} < \lambda$), the sequence is considered random (or non-random), where $\lambda$ is a predefined threshold known as the \emph{significance level}. For cryptographic purposes, $\lambda$ is usually set between 0.001 and 0.01~\cite{rukhin2001statistical}. Moreover, the proportion range of $(1-\lambda) \pm 3 \sqrt{\lambda (1-\lambda)/N}$ is considered acceptable, where $N \geq 1/ \lambda$ indicates the sample size (number of sequences). The NIST suite and its corresponding parameters are listed in~\ref{tab:NIST_parameters}, while a brief explanation of the suite can be found in~\cite{rukhin2001statistical}.\\

To ensure the randomness of the ECGA, we tested numerous optimized random sequences generated by our generator using all the tests in the NIST 800-22 test suite. For the experiments, we used a significance level $\lambda = 0.01$, a sample size $N \in \lbrace 100,800 \rbrace$, and a sequence of length $n = 10^{6}$ bits.
We converted 100 resultant sequences generated by the ECGA based on the randomly selected parameters listed 
in~\ref{tab:parameters} into their binary representation for the NIST analysis. Each generated sequence values lie in the range of $[0, 2^{8}-1]$, resulting in a total of $(8 \times 100\times 10^{6})$ bits for the evaluation of the NIST test suite.
We performed NIST tests on two data samples: firstly, by taking the first $(1 \times 10^{8})$ bits from the total of $(8 \times 10^{8})$ bits with $N = 100$, and then by taking the total $(8 \times 10^{8})$ bits with $N = 800$. The results are presented 
in~\ref{tab:NIST_Results}.
The results indicate that for $N=100$, $p_{\text{value}} \geq 0.01$ for all tests except the block frequency test, for which $p_{\text{value}}= 0.006$, which is very close to the acceptable value of $0.01$. Additionally, for $N=100$, the proportion of all the tests is greater than the lower bound of acceptable proportion, which is $0.96$.
Moreover, for $N=800$, $p_{\text{value}} \geq 0.01$ and $\text{proportion} \geq 0.97$ for each test included in the NIST test suite, where $0.97$ is the lower bound of the acceptable proportion for a sample size of 800. As a result, the ECGA passed all tests and is capable of generating highly random sequences.\\

We compared the results of the ECGA with the state-of-the-art generators~\cite{tutueva2020adaptive, talhaoui2021new, paul2021design, hamza2017novel, haider2022novel} using the NIST test suite, the results are presented 
in~\ref{tab:NIST_Comp_1},~\ref{tab:NIST_Comp_2}, and~\ref{tab:NIST_Comp_3} and are summarized as follows.
\begin{enumerate}[itemsep=0pt]
\item[1)] The results listed in~\ref{tab:NIST_Comp_1} are based on 100 different random sequences each of length $10^{6}$ bits. The ECGA passed all 100 sequences, attaining a proportion of 1 for six different tests. On the other hand, generators~\cite{tutueva2020adaptive},~\cite{talhaoui2021new}, and~\cite{paul2021design} attained proportions of 1 for three, five, and two tests, respectively. The ECGA also satisfied the acceptable proportion value $ 0.96$ for all the tests listed in~\ref{tab:NIST_Comp_1}. However, the generator~\cite{talhaoui2021new} failed the Random Excursions Test (RET)  with a proportion of 0.72. The main objective of RET is to analyze the occurrence of a specific number of visits in a cumulative sum random walk. The purpose of conducting this test is to assess whether the number of visits to a particular state within a cycle deviates from the expected frequency for a random sequence. The test comprises a total of eight individual tests, each focusing on one specific state: $-4, -3, -2, -1, +1, +2, +3,$ and $+4$. 
Thus, the ECGA not only passed all the tests but also attained the highest proportion for the maximum number of tests compared to the generators in~\cite{tutueva2020adaptive, talhaoui2021new, paul2021design}.
\item[2)] \ref{tab:NIST_Comp_2} compares the results based on eight distinct sequences, each of length $10^{6}$ bits. The ECGA passed the $p_{\text{value}}$ criterion for each test, while generator~\cite{hamza2017novel} failed the Non-Overlapping Template Matching Test with a $p_{\text{value}}$ of 0. Furthermore, our generator achieved $p_{\text{value}} \geq 0.9$ for five tests, compared to only one test for generator~\cite{hamza2017novel}. Hence, our generator outperforms~\cite{hamza2017novel}.
\item[3)] The comparison results of 30 random sequences with a total of $(30 \times 10^{6})$ bits are illustrated 
in~\ref{tab:NIST_Comp_3}. Out of 41 tests, the ECGA attained a proportion value of 1 for 37 tests, while 
the generator~\cite{haider2022novel} attained a proportion of 1 against 30 tests. Thus,  the ECGA passed more tests with a proportion rate of 1 when compared with the generator~\cite{haider2022novel}.
\end{enumerate}

Thus, the ECGA performed better than other existing PRNGs based on the NIST statistical test suite. 
The ECGA passed all the tests with the highest proportion rate for the maximum number of tests when compared to the generators~\cite{tutueva2020adaptive, talhaoui2021new, paul2021design, hamza2017novel, haider2022novel}. Therefore, it can be concluded that the ECGA is suitable for various applications that require randomness and unpredictability.

\begin{table*}[!t]
 \setlength{\tabcolsep}{-2.5pt}   
      \caption{Results of the NIST tests.}
        \label{tab:NIST_Results}
   \begin{tabular*}{\textwidth}{l @{\extracolsep{\fill}}  cccccc}
       \toprule  
   Test     & \multicolumn{3}{c}{$N=100$} & \multicolumn{3}{c}{$N=800$}  \\
 & \multicolumn{3}{c}{$\text{Total bits} = 1 \times 10^{8}$} & \multicolumn{3}{c}{$\text{Total bits} = 8 \times 10^{8}$}  \\    \cline{2-4}\cline{5-7} 
      
  	&	       $P_{\text{value}}$ 	&	Proportion	&	       Result  &	       $P_{\text{value}}$ 	&	Proportion	&	       Result	      \\ \midrule
Frequency Monobit       	&	0.6371	&	1	&	       Pass    	&	0.9835	&	0.99	&	       Pass    \\
Block Frequency~$(M=128)$        	&	0.0062	&	1	&	       Fail    	&	0.0951	&	0.99	&	       Pass    \\
Cusum Forward   	&	0.0757	&	1	&	       Pass    	&	0.7019	&	0.99	&	       Pass    \\
Cusum Reverse   	&	0.7399	&	1	&	       Pass    	&	0.8794	&	1.00	&	       Pass    \\
Runs    	&	0.1626	&	1	&	       Pass    	&	0.1828	&	0.99	&	       Pass    \\
Longest Runs    	&	0.6163	&	1	&	       Pass    	&	0.9047	&	0.99	&	       Pass    \\
Rank    	&	0.1296	&	0.98	&	       Pass    	&	0.9781	&	0.98	&	       Pass    \\
Discrete Fourier Transform      	&	0.9463	&	0.98	&	       Pass    	&	0.6916	&	0.99	&	       Pass    \\
Non-overlapping Template~$(m=9)$      &	0.9558	&	1	&	       Pass    	&	0.9114	&	0.99	&	       Pass    \\
Overlapping Template~$(m=9)$    	&	0.0712	&	0.99	&	       Pass    	&	0.3023	&	0.99	&	       Pass    \\
Universal Test  	&	0.0966	&	0.99	&	       Pass    	&	0.2477	&	0.99	&	       Pass    \\
Approximate Entropy~$(m=10)$     	&	0.0308	&	0.99	&	       Pass    	&	0.8576	&	0.99	&	       Pass    \\
RET(x=-4)       	&	0.0635	&	0.97	&	       Pass    	&	0.3185	&	0.99	&	       Pass    \\
RET(x=-3)       	&	0.5510	&	0.98	&	       Pass    	&	0.1654	&	0.99	&	       Pass    \\
RET(x=-2)       	&	0.1165	&	0.98	&	       Pass    	&	0.7993	&	0.99	&	       Pass    \\
RET(x=-1)       	&	0.8195	&	1	&	       Pass    	&	0.2348	&	0.99	&	       Pass    \\
RET(x=1)        	&	0.6890	&	0.98	&	       Pass    	&	0.1654	&	0.98	&	       Pass    \\
RET(x=2)        	&	0.1284	&	1	&	       Pass    	&	0.2633	&	0.99	&	       Pass    \\
RET(x=3)        	&	0.5510	&	0.95	&	       Pass    	&	0.8143	&	0.98	&	       Pass    \\
RET(x=4)        	&	0.0333	&	0.98	&	       Pass    	&	0.1180	&	0.98	&	       Pass    \\
REVT(x=-9)      	&	0.0213	&	1	&	       Pass    	&	0.6562	&	0.99	&	       Pass    \\
REVT(x=-8)      	&	0.1284	&	1	&	       Pass    	&	0.7800	&	0.99	&	       Pass    \\
REVT(x=-7)      	&	0.4220	&	1	&	       Pass    	&	0.6138	&	0.99	&	       Pass    \\
REVT(x=-6)      	&	0.9411	&	1	&	       Pass    	&	0.7681	&	0.99	&	       Pass    \\
REVT(x=-5)      	&	0.5174	&	1	&	       Pass    	&	0.2715	&	0.99	&	       Pass    \\
REVT(x=-4)      	&	0.3372	&	1	&	       Pass    	&	0.5590	&	0.99	&	       Pass    \\
REVT(x=-3)      	&	0.8195	&	1	&	       Pass    	&	0.6392	&	0.99	&	       Pass    \\
REVT(x=-2)      	&	0.2645	&	1	&	       Pass    	&	0.6605	&	0.99	&	       Pass    \\
REVT(x=-1)      	&	0.5174	&	1	&	       Pass    	&	0.5300	&	0.99	&	       Pass    \\
REVT(x=1)       	&	0.6890	&	1	&	       Pass    	&	0.5968	&	0.99	&	       Pass    \\
REVT(x=2)               	&	0.1056	&	0.98	&	       Pass    	&	0.1617	&	0.99	&	       Pass    \\
REVT(x=3)       	&	0.7231	&	0.97	&	       Pass    	&	0.2373	&	0.98	&	       Pass    \\
REVT(x=4)       	&	0.2041	&	0.97	&	       Pass    	&	0.2043	&	0.98	&	       Pass    \\
REVT(x=5)       	&	0.1703	&	0.98	&	       Pass    	&	0.6690	&	0.97	&	       Pass    \\
REVT(x=6)       	&	0.8195	&	0.98	&	       Pass    	&	0.9547	&	0.98	&	       Pass    \\
REVT(x=7)       	&	0.8195	&	0.98	&	       Pass    	&	0.2527	&	0.98	&	       Pass    \\
REVT(x=8)       	&	0.4528	&	0.97	&	       Pass    	&	0.9193	&	0.98	&	       Pass    \\
REVT(x=9)       	&	0.3641	&	0.97	&	       Pass    	&	0.5968	&	0.98	&	       Pass    \\
Serial 1~$(m=16)$        	&	0.4373	&	0.98	&	       Pass    	&	0.1956	&	0.99	&	       Pass    \\
Serial 2~$(m=16)$        	&	0.0329	&	0.98	&	       Pass    	&	0.2606	&	0.99	&	       Pass    \\
Linear Complexity~$(M=500)$       	&	0.1719	&	0.99	&	       Pass    	&	0.5930	&	0.99	&	       Pass    \\

 \bottomrule
\end{tabular*}
 \end{table*}

 \begin{table}[!t]
      \caption{Comparison of the NIST results corresponding to 100 sequences each of length $10^{6}$ bits.}
        \label{tab:NIST_Comp_1}
   \begin{tabular*}{\textwidth}{l @{\extracolsep{\fill}}  c c cc }
       \toprule  
 Test  & \multicolumn{4}{c}{Proportion} \\ \cline{2-5}
   	&	ECGA	&	Ref.~\cite{tutueva2020adaptive}	&	 Ref.~\cite{talhaoui2021new}  &
   	Ref.~\cite{paul2021design}     	\\\midrule
FT	&	1	&	0.99	&	0.98	&	0.99	\\
BFT	&	1	&	0.98	&	0.99	&	1	\\
CST	&	1	&	0.99	&	1	&	0.99	\\
RT	&	1	&	0.99	&	1	&	0.96	\\
LRT    	&	1	&	1	&	0.99	&	1	\\
RRT	&	0.98	&	1	&	1	&	0.99	\\
DFT      	&	0.98	&	0.98	&	0.96	&	0.99	\\
NTMT	&	1	&	0.99	&	0.93	&	0.99	\\
OTMT	&	0.99	&	0.99	&	0.89	&	0.99	\\
UT	&	0.99	&	0.98	&	0.99	&	0.99	\\
AET	&	0.99	&	0.98	&	0.99	&	0.98	\\
RET	&	0.98	&	0.98	&	0.72	&	0.98	\\
REVT	&	0.99	&	1	&	0.97	&	0.98	\\
ST 	&	0.98	&	0.99	&	1	&	0.99	\\
LCT	&	0.99	&	0.99	&	1	&	0.99	\\ 
  \bottomrule
\end{tabular*}
\\
      \caption{Comparison of the NIST results for $8 \times 10^{6}$ bits.}
        \label{tab:NIST_Comp_2}
  \setlength\extrarowheight{-4.5pt}
   \begin{tabular*}{\textwidth}{l @{\extracolsep{\fill}}  cc}
       \toprule  
 Test  & \multicolumn{2}{c}{$P_{\text{value}}$} \\ \cline{2-3}
   	&	ECGA	&	Ref.~\cite{hamza2017novel}	   	\\\midrule
FT	&	0.9032	&	0.1428	\\
BFT	&	0.0596	&	0.2028	\\
CST	&	0.9180	&	0.2666	\\
RT	&	0.9093	&	0.2859	\\
LRT    	&	0.0118	&	0.8303	\\
RRT	&	0.0751	&	0.8813	\\
DFT      	&	0.9689	&	0.1729	\\
NTMT	&	0.5027	&	0.0000	\\
OTMT	&	0.8779	&	0.5782	\\
UT	&	0.1660	&	0.6342	\\
AET	&	0.9693	&	0.0216	\\
RET	&	0.6341	&	0.7147	\\
REVT	&	0.4663	&	0.5142	\\
ST 1	&	0.0157	&	0.3608	\\
ST 2	&	0.0280	&	0.4480	\\
LCT	&	0.1999	&	0.9511	\\

  \bottomrule
\end{tabular*}
 \end{table}
 \begin{table}[!t]
      \caption{Comparison of the NIST results corresponding to 30 sequences each of length $10^{6}$ bits.}
        \label{tab:NIST_Comp_3}
   \begin{tabular*}{\textwidth}{l @{\extracolsep{\fill}}  cc}
       \toprule  
 Test  & \multicolumn{2}{c}{Proportion} \\ \cline{2-3}
   	&	ECGA	&	Ref.~\cite{haider2022novel}  	\\\midrule

FT	&	1	&	0.9667	\\
BFT	&	1	&	1	\\
CST (F)	&	1	&	0.9667	\\
CST(R)  	&	1	&	0.9667	\\
RT	&	1	&	0.9667	\\
LRT    	&	1	&	0.9667	\\
RRT	&	1	&	1	\\
DFT      	&	1	&	1	\\
NTMT	&	1	&	1	\\
OTMT	&	1	&	1	\\
UT	&	1	&	1	\\
AET	&	1	&	1	\\
RET(x=-4)       	&	1	&	1	\\
RET(x=-3)       	&	1	&	1	\\
RET(x=-2)       	&	1	&	1	\\
RET(x=-1)       	&	1	&	0.9643	\\
RET(x=1)        	&	0.9375	&	1	\\
RET(x=2)        	&	1	&	0.9643	\\
RET(x=3)        	&	1	&	0.9643	\\
RET(x=4)        	&	1	&	0.9643	\\
REVT(x=-9)      	&	1	&	1	\\
REVT(x=-8)      	&	1	&	1	\\
REVT(x=-7)      	&	1	&	1	\\
REVT(x=-6)      	&	1	&	1	\\
REVT(x=-5)      	&	1	&	1	\\
REVT(x=-4)      	&	1	&	1	\\
REVT(x=-3)      	&	1	&	1	\\
REVT(x=-2)      	&	1	&	1	\\
REVT(x=-1)      	&	1	&	1	\\
REVT(x=1)       	&	1	&	1	\\
REVT(x=2)               	&	1	&	1	\\
REVT(x=3)       	&	1	&	1	\\
REVT(x=4)       	&	1	&	1	\\
REVT(x=5)       	&	1	&	1	\\
REVT(x=6)       	&	1	&	0.9643	\\
REVT(x=7)       	&	1	&	1	\\
REVT(x=8)       	&	0.9375	&	1	\\
REVT(x=9)       	&	0.9375	&	1	\\
Serial 1        	&	0.9667	&	1	\\
Serial 2        	&	1	&	1	\\
LCT	&	1	&	0.9667	\\

\bottomrule
\end{tabular*}
 \end{table}
\subsection{Entropy analysis}
The degree of uncertainty in a PRNG can be measured by its information 
entropy~\cite{shannon1948mathematical}, which is typically expressed in bits. 
A good PRNG should produce sequences with high entropy, meaning a greater degree of uncertainty. 

In this study, 100 sequences of length $10^{6}$ with values ranging from 0 to $2^{8}-1$ were generated and their entropy was calculated. The optimal entropy denoted by $H_{\text{max}}$, in our case, is 8. The average entropy of the 100 sequences before optimization was between 6.7702 and 7.1084, with an average of 6.9305. However, after optimization, all 100 sequences achieved their maximum entropy of 8, as demonstrated in~\ref{tab:Exp_A} and~\ref{fig:Entropy}. These findings indicate that the ECGA can be very useful for cryptographic purposes, as it significantly increased the entropy of the generated sequences. Furthermore, the results listed in~\ref{tab:Entropy_Comp} demonstrates that  the ECGA outperformed several 
state-of-the-art PRNGs~\cite{zang2022construction, shi2021hybrid, agarwal2021designing, barani2020new, zhao2019self, wang2019pseudo} in generating sequences with optimal entropy.
\subsection{Period analysis}
To ensure that a PRNG generates a sequence that is sufficiently random and has a long enough period for its intended use, it is crucial to perform the period test~\cite{kalos2009monte}  on it.
The period is a significant attribute of a sequence generated by a PRNG, as it indicates the length of the shortest repeating cycle present in the sequence, if it exists. Specifically, it is the smallest positive integer $T$ for which the $k$-th element of the sequence matches the $(k+T)$-th element for all $k \geq 0$. If the period of a sequence is equal to its length, then it is the optimal period of the sequence denoted by $T_{\text{max}}$.\\

We evaluated the effectiveness of the ECGA by generating 100 random sequences, each containing $10^{6}$ numbers, using the ECGA. We analyzed the strength of the sequences by measuring their period, both before and after optimization. Our goal is to determine how much the period increased after optimization. The results are presented 
in~\ref{tab:Exp_A}. Our findings indicate that, before optimization, the sequences have periods ranging from 192 to 998712, while after optimization, all sequences have an optimal period of $10^{6}$. This suggests that our optimization algorithm has significantly increased the period of the random sequences, making them suitable for cryptographic applications. 
In other words, all the generated sequences have an optimal period and can be used for secure communication.
\subsection{Hurst exponent}
The Hurst exponent~\cite{hurst1951long} is a statistical test, that determines the trend in data. It is denoted as $\rm{H_E}$ and falls between 0 and 1. There are three possible scenarios when calculating $\rm{H_E}$:
\begin{enumerate}[itemsep=0pt]
\item[1)]
If $\rm{H_E}=0.5$, the data is random or independent, meaning there is no correlation between the current and previous values.
\item[2)] If $0.5<\rm{H_E}\leq1$, the data is persistent, meaning if there is an increasing trend in the values, the next values will likely follow the increasing trend.
\item[3)] If $0\leq \rm{H_E}<0.5$, the data is anti-persistent, meaning if there is an increasing trend in the values, the next values will likely follow the decreasing trend. 
\end{enumerate}
A value closer to 0.5 indicates more random data. A good PRNG should have an $\rm{H_E}$ value close to 0.5. The rescaled range (R/S) analysis~\cite{gilmore2002investigation} is the most commonly used method to compute $\rm{H_E}$.

We have calculated the Hurst exponent ($\rm{H_E}$)  using the (R/S) method for 100 sequences, each of length $10^6$. The results are presented in~\ref{tab:Exp_B} and shown in~\ref{fig:HExp}. The results indicate that before optimization, $\rm{H_E}$ ranged from 0.0665 to 0.2900, while after optimization, it ranged from 0.4916 to 0.5410. This demonstrates that after optimization, all generated sequences have $\rm{H_E}$ values very close to the ideal value of 0.5, indicating that the sequences are highly random. We have also shown the Hurst plots of the first five sequences in~\ref{fig:HE}.

We also compared the Hurst exponent of the ECGA with the state-of-the-art 
generators~\cite{xia2018novel, gayoso2013pseudorandom, gayoso2019general}. 
The results are presented in~\ref{tab:HE_Comp}, which shows that the ECGA $\rm{H_E}$ values are closer to the ideal value of 0.5 compared to the  
generators~\cite{xia2018novel, gayoso2013pseudorandom, gayoso2019general}. Thus, the ECGA can generate highly random sequences when compared with the generators described in~\cite{xia2018novel, gayoso2013pseudorandom, gayoso2019general}. 
\begin{table*}
 \setlength{\tabcolsep}{-3.5pt}   
      \caption{Results of entropy and period of the 100 randomly generated sequences by the ECGA. }
        \label{tab:Exp_A}
 \setlength\extrarowheight{-3.5pt}
   \begin{tabular*}{\textwidth}{l @{\extracolsep{\fill}} ccccccccccccc}
       \toprule  
Seq.  	&	$H(\Omega_{I})$	&	$H(\Omega_{Z})$	&	$H_{\text{max}}$	&	$T(\Omega_{I})$	&	$T(\Omega_{Z})$	&	
$T_{\text{max}}$	&	Seq. 	&	$H(\Omega_{I})$	&	$H(\Omega_{Z})$	&	$H_{\text{max}}$	&	$T(\Omega_{I})$	&	$T(\Omega_{Z})$	&	
$T_{\text{max}}$	\\
No. & &&&& && No. &&& &\\ \midrule
1	&	6.8394	&	8.000	&	8	&	998712	&	$10^{6}$	&	$10^{6}$	&	51	&	6.9321	&	8.000	&	8	&	998712	&	$10^{6}$	&	$10^{6}$	\\
2	&	6.9898	&	8.000	&	8	&	998712	&	$10^{6}$	&	$10^{6}$	&	52	&	6.9645	&	8.000	&	8	&	998712	&	$10^{6}$	&	$10^{6}$	\\
3	&	6.9723	&	8.000	&	8	&	998712	&	$10^{6}$	&	$10^{6}$	&	53	&	6.9934	&	8.000	&	8	&	998712	&	$10^{6}$	&	$10^{6}$	\\
4	&	6.9582	&	8.000	&	8	&	998712	&	$10^{6}$	&	$10^{6}$	&	54	&	6.8709	&	8.000	&	8	&	998712	&	$10^{6}$	&	$10^{6}$	\\
5	&	6.9979	&	8.000	&	8	&	998712	&	$10^{6}$	&	$10^{6}$	&	55	&	6.8391	&	8.000	&	8	&	998712	&	$10^{6}$	&	$10^{6}$	\\
6	&	6.9866	&	8.000	&	8	&	998712	&	$10^{6}$	&	$10^{6}$	&	56	&	6.8779	&	8.000	&	8	&	998712	&	$10^{6}$	&	$10^{6}$	\\
7	&	6.9683	&	8.000	&	8	&	998712	&	$10^{6}$	&	$10^{6}$	&	57	&	6.8565	&	8.000	&	8	&	998712	&	$10^{6}$	&	$10^{6}$	\\
8	&	6.9059	&	8.000	&	8	&	998712	&	$10^{6}$	&	$10^{6}$	&	58	&	6.8580	&	8.000	&	8	&	998712	&	$10^{6}$	&	$10^{6}$	\\
9	&	6.9618	&	8.000	&	8	&	998712	&	$10^{6}$	&	$10^{6}$	&	59	&	6.9623	&	8.000	&	8	&	998712	&	$10^{6}$	&	$10^{6}$	\\
10	&	6.9603	&	8.000	&	8	&	998712	&	$10^{6}$	&	$10^{6}$	&	60	&	6.9011	&	8.000	&	8	&	998712	&	$10^{6}$	&	$10^{6}$	\\
11	&	6.9735	&	8.000	&	8	&	998712	&	$10^{6}$	&	$10^{6}$	&	61	&	7.1084	&	8.000	&	8	&	192	&	$10^{6}$	&	$10^{6}$	\\
12	&	6.9603	&	8.000	&	8	&	998712	&	$10^{6}$	&	$10^{6}$	&	62	&	6.9376	&	8.000	&	8	&	192	&	$10^{6}$	&	$10^{6}$	\\
13	&	6.8841	&	8.000	&	8	&	192	&	$10^{6}$	&	$10^{6}$	&	63	&	6.8836	&	8.000	&	8	&	192	&	$10^{6}$	&	$10^{6}$	\\
14	&	6.9702	&	8.000	&	8	&	192	&	$10^{6}$	&	$10^{6}$	&	64	&	6.8466	&	8.000	&	8	&	192	&	$10^{6}$	&	$10^{6}$	\\
15	&	6.9466	&	8.000	&	8	&	192	&	$10^{6}$	&	$10^{6}$	&	65	&	6.9468	&	8.000	&	8	&	192	&	$10^{6}$	&	$10^{6}$	\\
16	&	6.9702	&	8.000	&	8	&	192	&	$10^{6}$	&	$10^{6}$	&	66	&	6.8697	&	8.000	&	8	&	192	&	$10^{6}$	&	$10^{6}$	\\
17	&	6.9012	&	8.000	&	8	&	192	&	$10^{6}$	&	$10^{6}$	&	67	&	6.9697	&	8.000	&	8	&	192	&	$10^{6}$	&	$10^{6}$	\\
18	&	6.8672	&	8.000	&	8	&	192	&	$10^{6}$	&	$10^{6}$	&	68	&	6.9201	&	8.000	&	8	&	192	&	$10^{6}$	&	$10^{6}$	\\
19	&	6.8973	&	8.000	&	8	&	192	&	$10^{6}$	&	$10^{6}$	&	69	&	7.0062	&	8.000	&	8	&	192	&	$10^{6}$	&	$10^{6}$	\\
20	&	6.9167	&	8.000	&	8	&	192	&	$10^{6}$	&	$10^{6}$	&	70	&	6.8973	&	8.000	&	8	&	192	&	$10^{6}$	&	$10^{6}$	\\
21	&	6.9350	&	8.000	&	8	&	192	&	$10^{6}$	&	$10^{6}$	&	71	&	6.9167	&	8.000	&	8	&	192	&	$10^{6}$	&	$10^{6}$	\\
22	&	6.9377	&	8.000	&	8	&	192	&	$10^{6}$	&	$10^{6}$	&	72	&	6.9519	&	8.000	&	8	&	192	&	$10^{6}$	&	$10^{6}$	\\
23	&	6.8711	&	8.000	&	8	&	192	&	$10^{6}$	&	$10^{6}$	&	73	&	6.9070	&	8.000	&	8	&	998712	&	$10^{6}$	&	$10^{6}$	\\
24	&	7.0470	&	8.000	&	8	&	192	&	$10^{6}$	&	$10^{6}$	&	74	&	6.8561	&	8.000	&	8	&	998712	&	$10^{6}$	&	$10^{6}$	\\
25	&	6.9337	&	8.000	&	8	&	998712	&	$10^{6}$	&	$10^{6}$	&	75	&	6.9382	&	8.000	&	8	&	998712	&	$10^{6}$	&	$10^{6}$	\\
26	&	7.0381	&	8.000	&	8	&	998712	&	$10^{6}$	&	$10^{6}$	&	76	&	6.8256	&	8.000	&	8	&	998712	&	$10^{6}$	&	$10^{6}$	\\
27	&	6.7702	&	8.000	&	8	&	998712	&	$10^{6}$	&	$10^{6}$	&	77	&	7.0330	&	8.000	&	8	&	998712	&	$10^{6}$	&	$10^{6}$	\\
28	&	6.9531	&	8.000	&	8	&	998712	&	$10^{6}$	&	$10^{6}$	&	78	&	6.9558	&	8.000	&	8	&	998712	&	$10^{6}$	&	$10^{6}$	\\
29	&	6.9798	&	8.000	&	8	&	998712	&	$10^{6}$	&	$10^{6}$	&	79	&	6.8590	&	8.000	&	8	&	998712	&	$10^{6}$	&	$10^{6}$	\\
30	&	6.9849	&	8.000	&	8	&	998712	&	$10^{6}$	&	$10^{6}$	&	80	&	6.9558	&	8.000	&	8	&	998712	&	$10^{6}$	&	$10^{6}$	\\
31	&	6.9171	&	8.000	&	8	&	998712	&	$10^{6}$	&	$10^{6}$	&	81	&	7.0436	&	8.000	&	8	&	998712	&	$10^{6}$	&	$10^{6}$	\\
32	&	6.8892	&	8.000	&	8	&	998712	&	$10^{6}$	&	$10^{6}$	&	82	&	7.0149	&	8.000	&	8	&	998712	&	$10^{6}$	&	$10^{6}$	\\
33	&	6.7820	&	8.000	&	8	&	998712	&	$10^{6}$	&	$10^{6}$	&	83	&	6.9573	&	8.000	&	8	&	998712	&	$10^{6}$	&	$10^{6}$	\\
34	&	6.9956	&	8.000	&	8	&	998712	&	$10^{6}$	&	$10^{6}$	&	84	&	6.9755	&	8.000	&	8	&	998712	&	$10^{6}$	&	$10^{6}$	\\
35	&	6.9344	&	8.000	&	8	&	998712	&	$10^{6}$	&	$10^{6}$	&	85	&	6.9102	&	8.000	&	8	&	192	&	$10^{6}$	&	$10^{6}$	\\
36	&	6.9914	&	8.000	&	8	&	998712	&	$10^{6}$	&	$10^{6}$	&	86	&	6.8308	&	8.000	&	8	&	192	&	$10^{6}$	&	$10^{6}$	\\
37	&	6.9167	&	8.000	&	8	&	192	&	$10^{6}$	&	$10^{6}$	&	87	&	6.9702	&	8.000	&	8	&	192	&	$10^{6}$	&	$10^{6}$	\\
38	&	6.8908	&	8.000	&	8	&	192	&	$10^{6}$	&	$10^{6}$	&	88	&	6.8414	&	8.000	&	8	&	192	&	$10^{6}$	&	$10^{6}$	\\
39	&	6.9053	&	8.000	&	8	&	192	&	$10^{6}$	&	$10^{6}$	&	89	&	6.8961	&	8.000	&	8	&	192	&	$10^{6}$	&	$10^{6}$	\\
40	&	6.8607	&	8.000	&	8	&	192	&	$10^{6}$	&	$10^{6}$	&	90	&	6.9975	&	8.000	&	8	&	192	&	$10^{6}$	&	$10^{6}$	\\
41	&	6.9246	&	8.000	&	8	&	192	&	$10^{6}$	&	$10^{6}$	&	91	&	6.7963	&	8.000	&	8	&	192	&	$10^{6}$	&	$10^{6}$	\\
42	&	6.9831	&	8.000	&	8	&	192	&	$10^{6}$	&	$10^{6}$	&	92	&	6.9007	&	8.000	&	8	&	192	&	$10^{6}$	&	$10^{6}$	\\
43	&	7.0223	&	8.000	&	8	&	192	&	$10^{6}$	&	$10^{6}$	&	93	&	7.0093	&	8.000	&	8	&	192	&	$10^{6}$	&	$10^{6}$	\\
44	&	7.0118	&	8.000	&	8	&	192	&	$10^{6}$	&	$10^{6}$	&	94	&	6.9324	&	8.000	&	8	&	192	&	$10^{6}$	&	$10^{6}$	\\
45	&	6.9375	&	8.000	&	8	&	192	&	$10^{6}$	&	$10^{6}$	&	95	&	6.9415	&	8.000	&	8	&	192	&	$10^{6}$	&	$10^{6}$	\\
46	&	6.8607	&	8.000	&	8	&	192	&	$10^{6}$	&	$10^{6}$	&	96	&	6.9162	&	8.000	&	8	&	192	&	$10^{6}$	&	$10^{6}$	\\
47	&	7.0132	&	8.000	&	8	&	192	&	$10^{6}$	&	$10^{6}$	&	97	&	6.9733	&	8.000	&	8	&	998712	&	$10^{6}$	&	$10^{6}$	\\
48	&	6.9063	&	8.000	&	8	&	192	&	$10^{6}$	&	$10^{6}$	&	98	&	6.9034	&	8.000	&	8	&	998712	&	$10^{6}$	&	$10^{6}$	\\
49	&	6.8834	&	8.000	&	8	&	998712	&	$10^{6}$	&	$10^{6}$	&	99	&	6.8545	&	8.000	&	8	&	998712	&	$10^{6}$	&	$10^{6}$	\\
50	&	6.8348	&	8.000	&	8	&	998712	&	$10^{6}$	&	$10^{6}$	&	100	&	7.0006	&	8.000	&	8	&	998712	&	$10^{6}$	&	$10^{6}$	\\
 \bottomrule
\end{tabular*}
\smallskip  
      \caption{Comparison based on entropy analysis.}
        \label{tab:Entropy_Comp}
  \setlength\extrarowheight{-2.5pt}
   \begin{tabular*}{\textwidth}{l @{\extracolsep{\fill}}  ccccccc}
       \toprule  
 Scheme  &  ECGA &  Ref.~\cite{zang2022construction} & Ref.~\cite{shi2021hybrid} & 
 Ref.~\cite{agarwal2021designing} & 
 Ref.~\cite{ barani2020new} & 
 Ref.~\cite{zhao2019self} & Ref.~\cite{wang2019pseudo} \\ \midrule
Entropy & 8.000 & 7.9980 &  7.9985 & 7.9864 &  7.9937 & 7.9896 & 7.9692 \\

  \bottomrule
\end{tabular*}
 \end{table*}
\begin{table}
      \caption{Results of Hurst exponent $\rm{H_E}$  of the 100 randomly generated sequences by the ECGA. }
        \label{tab:Exp_B}
   \begin{tabular*}{\textwidth}{l @{\extracolsep{\fill}} l l l l l}
       \toprule  
Seq. 	&	$\rm{H_E}(\Omega_{I})$	&	$\rm{H_E}(\Omega_{Z})$	&	Seq. 	&	$\rm{H_E}(\Omega_{I})$	&	$\rm{H_E}(\Omega_{Z})$\\
No.	&		&		&	No.	&		&	\\\midrule
1	&	0.268	&	0.540	&	51	&	0.163	&	0.521		\\			
2	&	0.233	&	0.518	&	52	&	0.248	&	0.520		\\			
3	&	0.250	&	0.520	&	53	&	0.290	&	0.518		\\			
4	&	0.198	&	0.500	&	54	&	0.226	&	0.527		\\			
5	&	0.186	&	0.517	&	55	&	0.126	&	0.520		\\			
6	&	0.234	&	0.502	&	56	&	0.157	&	0.517		\\			
7	&	0.104	&	0.516	&	57	&	0.176	&	0.514		\\			
8	&	0.218	&	0.509	&	58	&	0.099	&	0.528		\\			
9	&	0.244	&	0.534	&	59	&	0.131	&	0.521		\\			
10	&	0.247	&	0.514	&	60	&	0.125	&	0.512		\\			
11	&	0.196	&	0.515	&	61	&	0.146	&	0.508		\\			
12	&	0.187	&	0.511	&	62	&	0.119	&	0.522		\\			
13	&	0.114	&	0.505	&	63	&	0.118	&	0.531		\\			
14	&	0.090	&	0.532	&	64	&	0.130	&	0.529		\\			
15	&	0.114	&	0.509	&	65	&	0.141	&	0.535		\\			
16	&	0.067	&	0.516	&	66	&	0.110	&	0.525		\\			
17	&	0.091	&	0.524	&	67	&	0.124	&	0.513		\\			
18	&	0.089	&	0.518	&	68	&	0.123	&	0.508		\\			
19	&	0.130	&	0.509	&	69	&	0.107	&	0.496		\\			
20	&	0.087	&	0.493	&	70	&	0.095	&	0.520		\\			
21	&	0.067	&	0.514	&	71	&	0.080	&	0.508		\\			
22	&	0.107	&	0.529	&	72	&	0.108	&	0.506		\\			
23	&	0.088	&	0.520	&	73	&	0.111	&	0.504		\\			
24	&	0.105	&	0.531	&	74	&	0.194	&	0.518		\\			
25	&	0.125	&	0.524	&	75	&	0.140	&	0.525		\\			
26	&	0.132	&	0.493	&	76	&	0.139	&	0.505		\\			
27	&	0.165	&	0.519	&	77	&	0.105	&	0.516		\\			
28	&	0.166	&	0.492	&	78	&	0.263	&	0.522		\\			
29	&	0.162	&	0.514	&	79	&	0.246	&	0.510		\\			
30	&	0.226	&	0.500	&	80	&	0.132	&	0.497		\\			
31	&	0.175	&	0.518	&	81	&	0.213	&	0.514		\\			
32	&	0.239	&	0.527	&	82	&	0.149	&	0.506		\\			
33	&	0.115	&	0.526	&	83	&	0.143	&	0.537		\\			
34	&	0.144	&	0.525	&	84	&	0.230	&	0.493		\\			
35	&	0.234	&	0.511	&	85	&	0.082	&	0.516		\\			
36	&	0.143	&	0.508	&	86	&	0.107	&	0.515		\\			
37	&	0.099	&	0.510	&	87	&	0.115	&	0.507		\\			
38	&	0.091	&	0.493	&	88	&	0.094	&	0.523		\\			
39	&	0.095	&	0.504	&	89	&	0.111	&	0.531		\\			
40	&	0.115	&	0.512	&	90	&	0.078	&	0.516		\\			
41	&	0.114	&	0.516	&	91	&	0.088	&	0.504		\\			
42	&	0.094	&	0.522	&	92	&	0.124	&	0.503		\\			
43	&	0.112	&	0.511	&	93	&	0.123	&	0.521		\\			
44	&	0.118	&	0.527	&	94	&	0.106	&	0.533		\\			
45	&	0.108	&	0.526	&	95	&	0.097	&	0.541		\\			
46	&	0.123	&	0.530	&	96	&	0.117	&	0.527		\\			
47	&	0.107	&	0.529	&	97	&	0.172	&	0.493		\\			
48	&	0.124	&	0.520	&	98	&	0.088	&	0.504		\\			
49	&	0.175	&	0.510	&	99	&	0.131	&	0.517		\\			
50	&	0.213	&	0.524	&	100	&	0.155	&	0.520		\\			
																														
 \bottomrule
\end{tabular*}
 \end{table}

\setcounter{figure}{0}
  \begin{figure*}[!t]
	  \centering
            \begin{tikzpicture}
            \centering
			\begin{axis}[
		 width=\linewidth,
			 height=5cm,
				xlabel= Pseudo-random number sequence $(\Omega)$ ,
				ylabel= $H(\Omega)$ ,
				grid=both,
				xmin=1 ,xmax=100,
				ymin=6 ,ymax=8,
			ytick distance = 0.4,
				xtick distance = 5,
				yticklabel style = {font=\small},
                xticklabel style = {font=\small},
                legend cell align={left},
                label style = {font=\small},
				legend style={at={(1,0.95)},anchor=north east, font = \small}
				]
				
				\addplot[smooth,ultra thick,blue,solid] %
				table[x=X,y=H_AP]{Data1.txt};
				\addlegendentry{ $\Omega_{Z}$  };
				\addplot[smooth, ultra thick,blue,dashdotted] %
				table[x=X,y=H_BP]{Data1.txt};
				\addlegendentry{ $\Omega_{I}$};
			\end{axis}
			\end{tikzpicture}
			\caption{Entropy $(H)$ of 100  pseudo-random number sequences randomly generated by the ECGA.}
			\label{fig:Entropy} 
\bigskip
	  \centering
            \begin{tikzpicture}
            \centering
			\begin{axis}[
			width=\linewidth,
			 height=5cm,
				xlabel= Pseudo-random number sequence $(\Omega)$ ,
				ylabel= $\rm{H_E}(\Omega)$ ,
				grid=both,
				xmin=1 ,xmax=100,
				ymin= 0 ,ymax=1,
				xtick distance = 5,
				yticklabel style = {font=\small},
                xticklabel style = {font=\small},
                legend cell align={left},
                label style = {font=\small},
				legend style={at={(1,0.9)},anchor=north east, font = \small}
				]
				
				\addplot[smooth,ultra thick,blue,solid] %
				table[x=X,y=HE_AP]{Data1.txt};
				\addlegendentry{ $\Omega_{Z}$  };
				\addplot[smooth, ultra thick,blue,dashdotted] %
				table[x=X,y=HE_BP]{Data1.txt};
				\addlegendentry{ $\Omega_{I}$};
			\end{axis}
			\end{tikzpicture}
			\caption{Hurst exponent ($\rm{H_E}$) of 100  pseudo-random number sequences randomly generated by the ECGA.}
			\label{fig:HExp} 
	\end{figure*}  
 
\begin{figure*}[!t]
	  \centering
            \begin{tikzpicture}
            \centering
			\begin{axis}[
	     width=0.35\linewidth,
 title style={at={(0.3,0.8)},anchor=north,yshift=-0.1},
			 title = {$\rm{H_E} = 0.5403$},
				xlabel= $\text{log}(n)$ ,
				ylabel= log(R/S) ,
				grid=both,
				xtick distance = 4,
				yticklabel style = {font=\small},
                xticklabel style = {font=\small},
                legend cell align={left},
                label style = {font=\small},
				legend style={at={(1,1)},anchor=north west, font = \small}
				]
				
				\addplot [only marks] table {Data_HE.txt};
                \addplot [no markers, thick, red]
      table [y={create col/linear regression={y=H1}}] {Data_HE.txt};
			\end{axis}
			\end{tikzpicture}
 \begin{tikzpicture}
            \centering
			\begin{axis}[
		     width=0.35\linewidth,
 title style={at={(0.3,0.8)},anchor=north,yshift=-0.1},
			 title = {$\rm{H_E} = 0.5183$},
				xlabel= $\text{log}(n)$ ,
				ylabel= log(R/S) ,
				grid=both,
				xtick distance = 4,
				yticklabel style = {font=\small},
                xticklabel style = {font=\small},
                legend cell align={left},
                label style = {font=\small},
				legend style={at={(1,1)},anchor=north west, font = \small}
				]
				
				\addplot [only marks] table {Data_HE.txt};
                \addplot [no markers, thick, red]
      table [y={create col/linear regression={y=H2}}] {Data_HE.txt};
			\end{axis}
			\end{tikzpicture}
			 \begin{tikzpicture}
            \centering
			\begin{axis}[
		     width=0.35\linewidth,
 title style={at={(0.3,0.8)},anchor=north,yshift=-0.1},
			 title = {$\rm{H_E} = 0.4996$},
				xlabel= $\text{log}(n)$ ,
				ylabel= log(R/S) ,
				grid=both,
				xtick distance = 4,
				yticklabel style = {font=\small},
                xticklabel style = {font=\small},
                legend cell align={left},
                label style = {font=\small},
				legend style={at={(1,1)},anchor=north west, font = \small}
				]
				
				\addplot [only marks] table {Data_HE.txt};
                \addplot [no markers, thick, red]
      table [y={create col/linear regression={y=H3}}] {Data_HE.txt};
			\end{axis}
			\end{tikzpicture}
\\~\\
			 \begin{tikzpicture}
            \centering
			\begin{axis}[
		     width=0.35\linewidth,
 title style={at={(0.3,0.8)},anchor=north,yshift=-0.1},
			 title = {$\rm{H_E} = 0.5054$},
				xlabel= $\text{log}(n)$ ,
				ylabel= log(R/S) ,
				grid=both,
				xtick distance = 4,
				yticklabel style = {font=\small},
                xticklabel style = {font=\small},
                legend cell align={left},
                label style = {font=\small},
				legend style={at={(1,1)},anchor=north west, font = \small}
				]
				
				\addplot [only marks] table {Data_HE.txt};
                \addplot [no markers, thick, red]
      table [y={create col/linear regression={y=H4}}] {Data_HE.txt};
			\end{axis}
			\end{tikzpicture}					
		 \begin{tikzpicture}
            \centering
			\begin{axis}[
		     width=0.35\linewidth,
            title style={at={(0.3,0.8)},anchor=north,yshift=-0.1},
			 title = {$\rm{H_E} = 0.5239$},
				xlabel= $\text{log}(n)$ ,
				ylabel= log(R/S) ,
				grid=both,
				xtick distance = 4,
				yticklabel style = {font=\small},
                xticklabel style = {font=\small},
                legend cell align={left},
                label style = {font=\small},
				legend style={at={(1,1)},anchor=north west, font = \small}
				]
				
				\addplot [only marks] table {Data_HE.txt};
                \addplot [no markers, thick, red]
      table [y={create col/linear regression={y=H5}}] {Data_HE.txt};
			\end{axis}
			\end{tikzpicture}		
			\caption{Hurst plot of  our fist five optimized pseudo-random sequences randomly generated by the ECGA.}
			\label{fig:HE} 
\end{figure*}
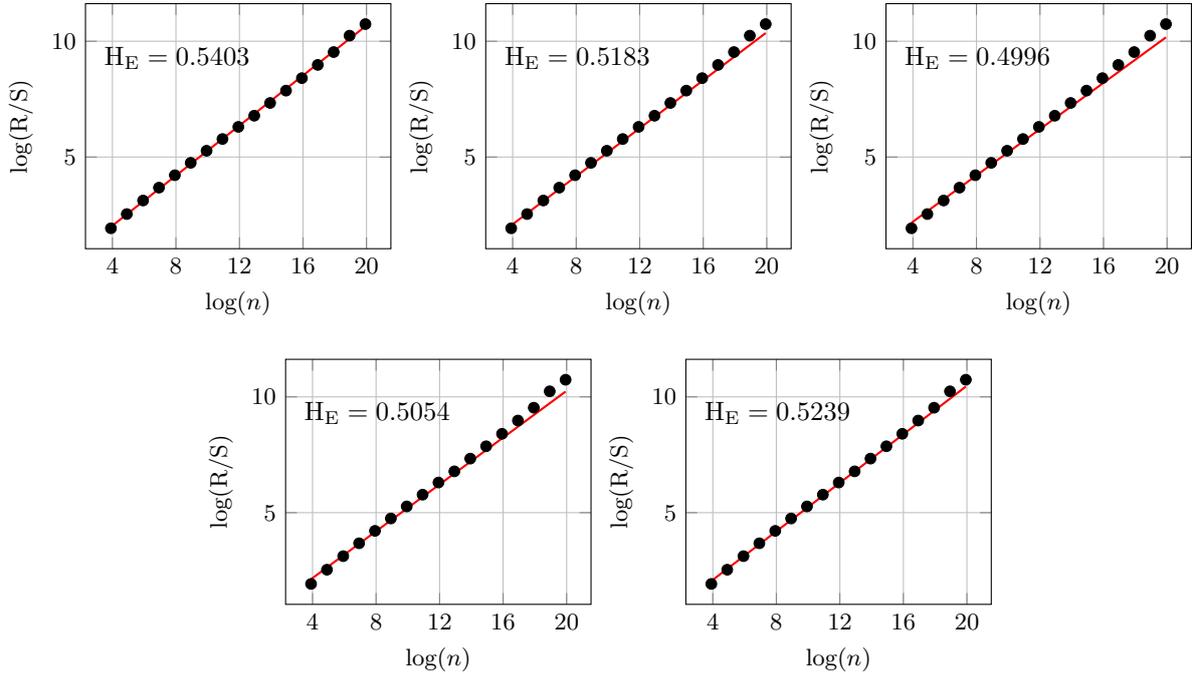  
\subsection{Correlation analysis}\label{sec_CC}
The correlation coefficient~\cite{pearson1901} is a crucial measure for determining the level of similarity between two random sequences of the same length. Given two random sequences, $\Omega=\lbrace \omega_{j} \rbrace_{j=1}^{l}$ and $\Omega'=\lbrace \omega_{j}' \rbrace_{j=1}^{l}$, with length $l$, the correlation coefficient $(R)$ is calculated using the following formula:

\begin{equation}
R(\Omega, \Omega') = \dfrac{\sum_{j=1}^{l}(\omega_{j}-\overline{\Omega})(\omega_{j}'-\overline{\Omega'})}{\sqrt{\sum_{j=1}^{l}(\omega_{j}-\overline{\Omega})^{2}}\sqrt{\sum_{j=1}^{l}(\omega_{j}'-\overline{\Omega'})^{2}}}.
\end{equation}

Here, $\overline{\Omega}$ and $\overline{\Omega'}$ represent the mean of $\Omega$ and $\Omega'$, respectively. 
The resulting $R(\Omega, \Omega')$ is a value between -1 and 1. When $ R(\Omega, \Omega')$ is close to 0, the sequences are considered independent, whereas a value of 1 or -1 indicates a high level of dependency.

We calculated $R(\Omega_{i}, \Omega_{j}')$ for 100 sequences we generated, where $i,j \in \lbrace 1, 2, ..., 100 \rbrace$ and $i \neq j$. From our experiments, we determined that the minimum and average values of the optimized generated sequences for all pairs of $i,j$ excluding when $i=j$ are 0 and 0.0638, respectively. Since the average value is very close to the ideal value of 0, we conclude that the ECGA can generate highly independent sequences.

\subsection{Key sensitivity analysis}
Key sensitivity analysis enables the study of how minor changes to the input parameters or initial conditions can cause changes in the resulting output.
If a PRNG has high sensitivity, even a slight change in the input can result in a significant difference in the output. 
This implies that a PRNG should exhibit a high level of sensitivity, even at the single-bit level~\cite{alvarez2006some}.

To demonstrate the high sensitivity of the ECGA, we conducted an experiment in which we slightly altered the parameters of  the ECGA.
For our purpose, we generated two sequences, $\Omega$ and $\Omega'$, which represent the original sequence and a slightly varied sequence, respectively, to analyze the sensitivity of the ECGA. 
A sequence $\Omega$ is generated using: the parameters of EC $E_{p,a,b }$ and the plain-image~(a) which are defined
in~\ref{tab:parameters}, $r \times s = 256 \times 256$, and $(\phi, \psi, \varphi) = (25,73,121)$.
Subsequently, slight modifications to $E_{p,a,b}$, $PI$, $r \times s$, and the triplet $(\phi,\psi,\varphi)$ resulted in the generation of four distinct sequences: $\Omega_{E_{p',a',b'}}'$, $\Omega_{PI'}'$, $\Omega_{r' \times s'}'$, and $\Omega_{(\phi', \psi', \varphi')}'$, respectively. The values of $E_{p',a',b'}$ are listed in~\ref{tab:parameters}, while $PI'$ represents the plain-image (b)  shown in~\ref{tab:parameters}. The dimensions of $r' \times s'$ were set to $512 \times 512$, and the triplet $(\phi',\psi',\varphi')$ was set to $(123,33,77)$.\\

We analyzed the sensitivity of the ECGA using three different methods:\\
1) by graphical representation;\\
2) by computing the number of bit change rate (NBCR); and\\
3) by computing the correlation coefficient.\\
\subsubsection{Graphical representation}
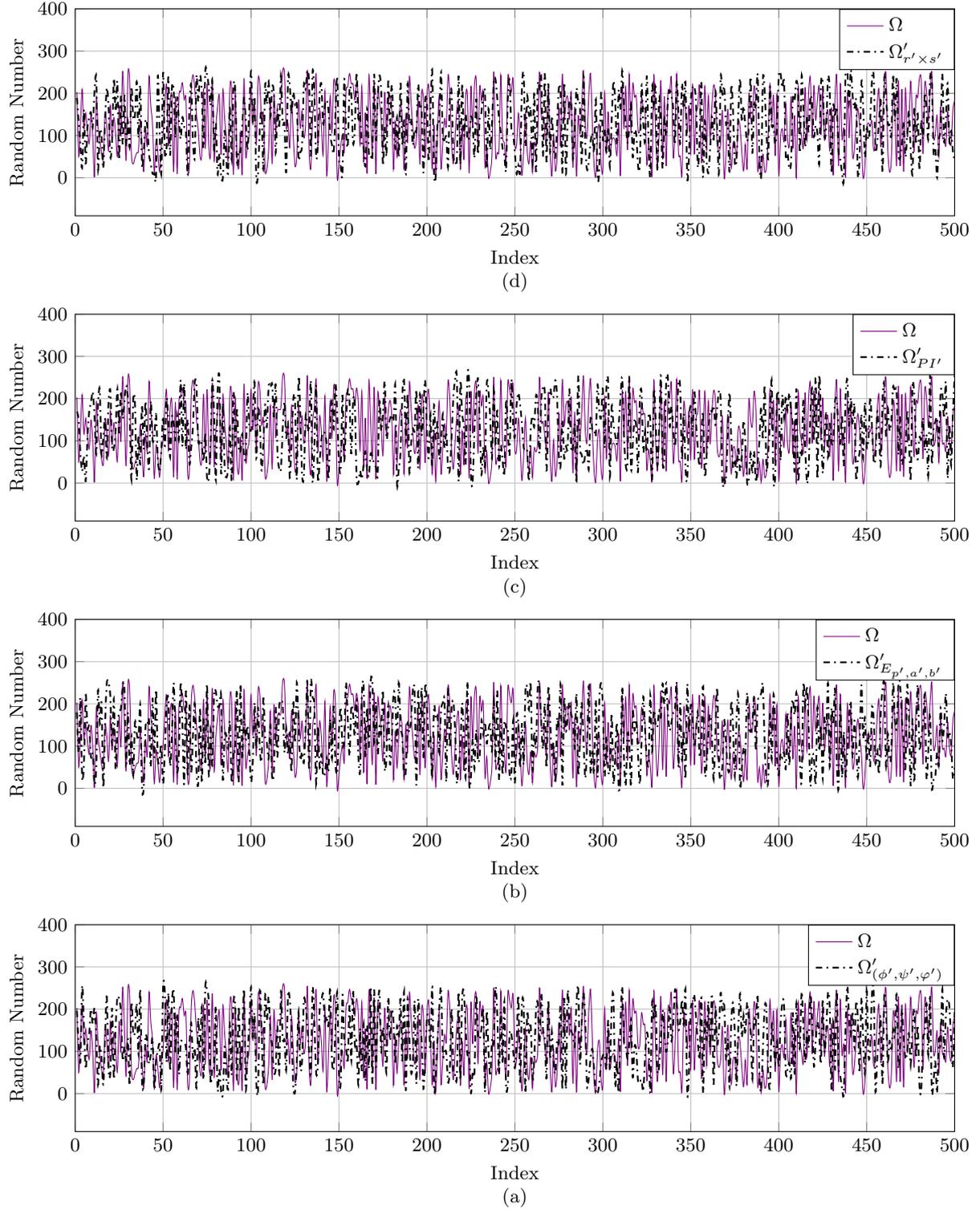
\begin{figure*}[!t]
	  \centering
            \begin{tikzpicture}
            \centering
			\begin{axis}[
		 width=\linewidth,
			 height=5cm,
			    xlabel style={align=center}, xlabel={Index\\(d)},
				ylabel= Random Number,
				grid=both,
				xmin=0 ,xmax=500,
				ymin=-90 ,ymax=400,
			ytick distance = 100,
				yticklabel style = {font=\small},
                xticklabel style = {font=\small},
                legend cell align={left},
                label style = {font=\small},
				legend style={at={(1,1)},anchor=north east, font = \small}
				]
				
				\addplot[smooth,thin,violet,solid] %
				table[x=X,y=S1]{Data_SA.txt};
				\addlegendentry{ $\Omega$  };
				\addplot[smooth, thick,black,dashdotted] %
				table[x=X,y=S2]{Data_SA.txt};
				\addlegendentry{ $\Omega_{r' \times s'}'$};
			\end{axis}
			\end{tikzpicture}
		\begin{tikzpicture}
            \centering
			\begin{axis}[
		 width=\linewidth,
			 height=5cm,
			    xlabel style={align=center}, xlabel={Index\\(c)},
				ylabel= Random Number,
				grid=both,
				xmin=0 ,xmax=500,
				ymin=-90 ,ymax=400,
			ytick distance = 100,
				yticklabel style = {font=\small},
                xticklabel style = {font=\small},
                legend cell align={left},
                label style = {font=\small},
				legend style={at={(1,1)},anchor=north east, font = \small}
				]
				
				\addplot[smooth,thin,violet,solid] %
				table[x=X,y=S1]{Data_SA.txt};
				\addlegendentry{ $\Omega$  };
				\addplot[smooth, thick,black,dashdotted] %
				table[x=X,y=S3]{Data_SA.txt};
				\addlegendentry{ $\Omega_{PI'}'$};
			\end{axis}
			\end{tikzpicture}	
	\begin{tikzpicture}
            \centering
			\begin{axis}[
		 width=\linewidth,
			 height=5cm,
			    xlabel style={align=center}, xlabel={Index\\(b)},
				ylabel= Random Number,
				grid=both,
				xmin=0 ,xmax=500,
				ymin=-90 ,ymax=400,
			ytick distance = 100,
				yticklabel style = {font=\small},
                xticklabel style = {font=\small},
                legend cell align={left},
                label style = {font=\small},
				legend style={at={(1,1)},anchor=north east, font = \small}
				]
				
				\addplot[smooth,thin,violet,solid] %
				table[x=X,y=S1]{Data_SA.txt};
				\addlegendentry{ $\Omega$  };
				\addplot[smooth, thick,black,dashdotted] %
				table[x=X,y=S4]{Data_SA.txt};
				\addlegendentry{ $\Omega_{E_{p',a',b'}}'$};
			\end{axis}
			\end{tikzpicture}
\begin{tikzpicture}
            \centering
			\begin{axis}[
		 width=\linewidth,
			 height=5cm,
			    xlabel style={align=center}, xlabel={Index\\(a)},
				ylabel= Random Number,
				grid=both,
				xmin=0 ,xmax=500,
				ymin=-90 ,ymax=400,
			ytick distance = 100,
				yticklabel style = {font=\small},
                xticklabel style = {font=\small},
                legend cell align={left},
                label style = {font=\small},
				legend style={at={(1,1)},anchor=north east, font = \small}
				]
				
				\addplot[smooth,thin,violet,solid] %
				table[x=X,y=S1]{Data_SA.txt};
				\addlegendentry{ $\Omega$  };
				\addplot[smooth, thick,black,dashdotted] %
				table[x=X,y=S5]{Data_SA.txt};
				\addlegendentry{$\Omega_{(\phi', \psi', \varphi')}'$};
			\end{axis}
			\end{tikzpicture}					
			\caption{Sensitivity of the ECGA corresponding to a change in parameters: (a) the triplet $(\phi', \psi', \varphi')$; (b) the elliptic curve $E_{p,a,b}$; (c) the plain-image $PI$; and (d) the dimension $r \times s$.}
			\label{fig:SA} 
	\end{figure*}

We analyzed the sensitivity of the ECGA  by visually displaying both the original sequence $\Omega$ and a slightly altered version of it, denoted as $\Omega'$.
The impact of various parameters is investigated and is presented in~\ref{fig:SA}. 
Specifically,~\ref{fig:SA}(a) depicts the effects of varying the parameters of the triplet $(\phi, \psi, \varphi)$, 
while~\ref{fig:SA}(b) shows the impact of changing the parameters of EC. 
\ref{fig:SA}(c) and~\ref{fig:SA}(d) demonstrate the effects of modifying the parameters $PI$ and the size $r \times s$, respectively. As illustrated in~\ref{fig:SA}, a slight modification in any of the parameters  $E_{p',a',b'}, PI', r' \times s',$ and $ (\phi', \psi', \varphi')$ resulted in a distinct sequence $\Omega'$ that differed from the original sequence $\Omega$. 
Hence, it can be concluded that the ECGA is highly sensitive to input parameters.\\

\subsubsection{Number of bit change rate}
The number of bit change rate (NBCR)~\cite{gentle2003random} is a common measure used to evaluate the sensitivity of a PRNG.  
To calculate NBCR for two randomly generated sequences $\Omega$ and $\Omega'$, we use the following equation:
\begin{equation}
\text{NBCR}(\Omega, \Omega')= \frac{d_{H}(\Omega,\Omega')}{n},
\end{equation}
where, $d_H$ denotes the Hamming distance between $\Omega$ and $\Omega'$, and $n$ represents the total number of bits in either sequence. An ideal value of NBCR is 50\%, which means that the closer the value of NBCR is to 50\%, the more sensitive the algorithm is.

We have computed the NBCR of the original sequence $\Omega$ and a slightly changed sequence $\Omega'$ and the results are listed in~\ref{tab:Corr_NBCR}. These results depict that the value of NBCR is very close to the optimal NBCR 50\%, which indicates that the ECGA is highly sensitive to the input parameters and thus applicable for security applications.

We also examined the NBCR of the proposed sequences both before and after the optimization process. 
The results, as shown in~\ref{tab:Corr_NBCR}, indicate that NBCR values are within the intervals of $[49.04, 51.19]$ and $[49.99, 50.04]$ for the sequences before and after optimization, respectively. Additionally, we compared the NBCR  of the ECGA with other PRNGs~\cite{yu2021design, agarwal2021designing, shi2021hybrid}, in terms of NBCR. 
\ref{tab:SA_comp} shows that the NBCR value of the ECGA is identical to the optimal value of $50\%$, while the NBCR of the other PRNGs is closer to $50\%$. Therefore, the ECGA is more sensitive to input parameters when compared 
to the generators~\cite{yu2021design, agarwal2021designing, shi2021hybrid}.

\begin{table*}[t!]  
      \caption{Comparison based on Hurst exponent $\rm{H_{E}}$ analysis.}
        \label{tab:HE_Comp}
   \begin{tabular*}{\textwidth}{l @{\extracolsep{\fill}}  cccc}
       \toprule  
 Scheme  &  ECGA & Ref.~\cite{xia2018novel} & Ref.~\cite{gayoso2013pseudorandom} & 
 Ref.~\cite{gayoso2019general}   \\ \midrule
 $\rm{H_{E}}$ & 0.5146 & 0.5261 & 0.5158 & 0.5241  \\

  \bottomrule
\end{tabular*}
\bigskip
      \caption{Correlation coefficient and NBCR between the original sequence $\Omega$ and a slightly varied sequence $\Omega'$.}
        \label{tab:Corr_NBCR}
\begin{tabular*}{\textwidth}{l @{\extracolsep{\fill}}  cccc }
       \toprule  
      
 Change in   &  \multicolumn{2}{c}{Before optimization}  & \multicolumn{2}{c}{After optimization} \\
 \cline{2-3}\cline{4-5}
 parameter  & $R(\Omega, \Omega')$  & $\text{NBCR}(\Omega, \Omega')$ & $R(\Omega, \Omega')$  & $\text{NBCR}(\Omega, \Omega')$\\
 \midrule
$r\times s$ & 0.0454 & 49.04 & -0.0017& 50.02\\
$PI$ & 0.1392 & 51.19 & 0.0015 & 49.99 \\
$E_{p,a,b}$ & 0.0005 & 50.00 & -0.0004 & 50.00\\
$\phi, \psi, \varphi $ & 0.0234 & 49.85 & -0.0016 & 50.04 \\ 
\bottomrule
\end{tabular*}
\bigskip
      \caption{Comparison of sensitivity analysis in terms of correlation coefficient and NBCR between the original sequence $\Omega$ and a slightly varied sequence $\Omega'$.}
        \label{tab:SA_comp}

\begin{tabular*}{\textwidth}{c @{\extracolsep{\fill}}  cccccc }
       \toprule  
      
  \multicolumn{3}{c}{$R(\Omega, \Omega')$}  & \multicolumn{4}{c}{$\text{NBCR}(\Omega, \Omega')$ 
  (\%)} \\
 \cline{1-3}\cline{4-7}
ECGA &  Ref.~\cite{agarwal2021designing} &  Ref.~\cite{zang2022construction} &
ECGA  & Ref.~\cite{yu2021design} &  Ref.~\cite{agarwal2021designing} 
&  Ref.~\cite{shi2021hybrid} \\ \midrule
0.0013 & 0.0016 & 0.0015 & 50.01 & 49.98 & 49.97 & 49.97 \\
\bottomrule
\end{tabular*}
\bigskip
 \setlength{\tabcolsep}{-2.5pt}   
      \caption{Comparison based on key space analysis.}
        \label{tab:Keyspace_Comp}
   \begin{tabular*}{\textwidth}{l @{\extracolsep{\fill}}  l l l l l l l l l}
       \toprule  
 Scheme  &  ECGA & Ref.~\cite{cang2021pseudo} & Ref.~\cite{agarwal2021designing} & 
 Ref.~\cite{barani2020new} &  Ref.~\cite{ayubi2020deterministic} &
 Ref.~\cite{zhao2019self} & Ref.~\cite{meranza2019pseudorandom} &
 Ref.~\cite{murillo2017novel} & Ref.~\cite{hamza2017novel} \\ \midrule
Key space & $2^{1280}$ & $2^{192}$ &  $2^{320}$ &  $2^{588}$ & $2^{232}$ & $2^{70}$ & $2^{222}$ & $2^{128}$ & $2^{279}$ \\

  \bottomrule
\end{tabular*}
 \end{table*}
 
\subsubsection{Correlation coefficient}
The sensitivity of the ECGA is also evaluated by calculating the correlation coefficient $R$ between two sequences, $\Omega$ and $\Omega'$, where $\Omega$ represents the original sequence while $\Omega'$ represents a slightly modified version of the original sequence. To ensure high sensitivity, $R(\Omega, \Omega')$ should be close to 0. We analyzed the results of $R(\Omega, \Omega_i')$ where $i$ belongs to the set $E_{p',a',b'}, PI', r' \times s', (\phi', \psi', \varphi')$, and presented 
in~\ref{tab:Corr_NBCR}. The results show that $R(\Omega, \Omega')$ values ranged between $[0.0005, 0.1392]$ and $[0.0004, 0.0017]$ before and after optimization, respectively. These results indicate that our designed generator is highly sensitive to its parameters, as both before and after optimization, the $R$ values are very close to 0. Furthermore, there is a significant improvement in the $R$ values after the optimization process.

Moreover, we conducted a comparison between the sensitivity of  the ECGA and the state-of-the-art 
generators~\cite{agarwal2021designing, zang2022construction}, using correlation coefficient as the metric. The outcomes of this comparison are shown in~\ref{tab:SA_comp}. The findings from the~\ref{tab:SA_comp} indicate that  the ECGA is more sensitive than the PRNGs~\cite{agarwal2021designing, zang2022construction}.

\subsection{Key space analysis}
The evaluation of the security of a cryptographic algorithm is closely linked to the concept of key 
space. When a PRNG is used for cryptographic purposes, it is essential to analyze its key space, which is the set of possible keys that can be used to generate a sequence. A larger key space makes the algorithm more resistant to exhaustive attacks, thereby improving its resistance to cryptanalysis.
To ensure the security of a cryptosystem, it is recommended that the key space should be at least 
$2^{128}$~\cite{alvarez2006some}. 

 The ECGA is based on several parameters, including $I$, $E_{p,a,b}$, $B_{z}$, $\phi$, $\psi$, and $\varphi$, as described 
in~\ref{sec:generator}. The SHA-256 hash code of $I$, $p$, $a$, and $b$ is also utilized. Additionally, the random sequence $B_{z}$ has at least $256$ bits, and the parameters $\phi$, $\psi$, and $\varphi$ range from 0 to 255.
As a result, the key space of the ECGA is at least $2^{256 \times 5}$. 
In comparison with the recommended key space of $2^{128}$, this is a very significant increase, indicating that  the ECGA is capable of withstanding modern cryptanalysis due to its extremely large key space.

We conducted a comparative analysis of the key space of the ECGA with that of the existing state-of-the-art 
generators~\cite{cang2021pseudo, agarwal2021designing, barani2020new, ayubi2020deterministic, zhao2019self, meranza2019pseudorandom, murillo2017novel, hamza2017novel}. 
The findings of our analysis are presented in~\ref{tab:Keyspace_Comp}. The results indicate that the ECGA has a superior key space in comparison to the PRNGs developed in~\cite{cang2021pseudo, agarwal2021designing, barani2020new, ayubi2020deterministic, zhao2019self, meranza2019pseudorandom, murillo2017novel, hamza2017novel}.
\section{Conclusion}\label{sec:Con}
We presented a novel method ECGA for the construction of an image-dependent pseudo-random number generator (IDPRNG) specifically for image-cryptographic applications. 
We addressed the limitations of traditional PRNGs by using a multi-objective genetic algorithm (MOGA) optimization method and integrating elliptic curves into our approach.
The ECGA comprises two key phases. During the initial phase, we utilize pixels from the image along with the parameters of the elliptic curve to generate an initial sequence of random numbers.
During the second phase, a genetic algorithm is utilized to enhance the generated sequence by maximizing a fitness function that is based on both the information entropy and period of the pseudo-random sequence.

We conducted thorough experiments and security evaluations to assess the performance of the ECGA. These evaluations covered a wide range of tests, including randomness analysis, entropy analysis, period analysis, correlation analysis, Hurst exponent analysis, key sensitivity analysis, and key space analysis.
Furthermore, we have compared the ECGA with the existing state-of-the-art generators, from which it is evident that the ECGA:
exhibits superior performance relative to other state-of-the-art 
generators~\cite{tutueva2020adaptive, talhaoui2021new, paul2021design, hamza2017novel, haider2022novel} as per the NIST statistical test suite;
outperformed several 
state-of-the-art generators~\cite{zang2022construction, shi2021hybrid, agarwal2021designing, barani2020new, zhao2019self, wang2019pseudo} in generating sequences with optimal entropy;
produces better  Hurst exponent results that are closer to the ideal value of 0.5  when compared with the  
generators~\cite{xia2018novel, gayoso2013pseudorandom, gayoso2019general};
is more sensitive to input parameters than the generators~\cite{yu2021design, agarwal2021designing, shi2021hybrid}; and
has a superior key space in comparison to the generators~\cite{cang2021pseudo, agarwal2021designing, barani2020new, ayubi2020deterministic, zhao2019self, meranza2019pseudorandom, murillo2017novel, hamza2017novel}.

Future work could explore further optimizations, evaluate the performance on large-scale data sets, and investigate the applicability of the IDPRNG in other domains requiring secure and unpredictable random number generation.\\

\bibliographystyle{plainnat}

\bibliography{bibfile}
 \end{document}